\documentclass[fleqn,10pt]{wlscirep}
\usepackage[utf8]{inputenc}
\usepackage[T1]{fontenc}
\usepackage[ruled]{algorithm2e}
\usepackage{braket}
\usepackage{subfig}
\usepackage{nameref}

\newcommand{\bs}[1]{\boldsymbol{#1}}
\newcommand{\nm}{\nonumber}

\title{Quantum topology optimization of ground structures using noisy intermediate-scale quantum devices}

\author[1,*]{Yuki Sato}
\author[1]{Ruho Kondo}
\author[1]{Satoshi Koide}
\author[1]{Seiji Kajita}
\affil[1]{Toyota Central R\&D Labs., Inc., 41-1, Yokomichi, Nagakute, Aichi 480-1192, Japan}

\affil[*]{yuki-sato@mosk.tytlabs.co.jp}
\begin{abstract}
To arrive at some viable product design, product development processes frequently use numerical simulations and mathematical programming techniques.
Topology optimization, in particular, is one of the most promising techniques for generating insightful design choices.
Topology optimization problems reduce to an NP-hard combinatorial optimization problem, where the combination of the existence or absence of the material at some positions is optimized.
In this study, we examine the usage of quantum computers as a potential solution to topology optimization problems.
The proposed method consists of two variational quantum algorithms (VQAs): the first solves the state equilibrium equation for all conceivable material configurations, while the second amplifies the likelihood of an optimal configuration in quantum superposition using the first VQA's quantum state.
Several experiments, including a real device experiment, show that the proposed method successfully obtained the optimal configurations.
These findings suggest that quantum computers could be a potential tool for solving topology optimization problems and they open the window to the near-future product designs.
\end{abstract}
\begin{document}

\flushbottom
\maketitle
\thispagestyle{empty}

\section*{Introduction}
Because of its potential to improve product development, topology optimization has been effectively employed in a variety of industries, including the automotive industry~\cite{Rozvany2009critical}.
The central notion of topology optimization is to replace structural optimization problems with a material distribution problem, which is a problem getting binary variable distributions with a value of $1$ represent the ``exist'' and $0$ the ``absence'' of the material.
Topology optimization is categorized into two types: the one is for continuum structures \cite{Rozvany2009critical} and the other for discrete structures, such as truss structures~\cite{Stolpe2016truss}.
For discrete structures, truss topology optimization methods were widely studied.
Ground structure methods, in which several pre-placed nodes connected by as many edges as feasible are prepared and then the existence or absence of each edge is optimized using the mixed integer programming, are commonly built in these methods~\cite{Kanno2010mixed, Yonekura2010global}.
It is well-known that determining minimum weight truss structures with discrete member sizes is an NP-hard problems~\cite{Stolpe2016truss}, and thus the heuristic-based approaches have been developed ~\cite{Kaveh2009particle, Sadollah2012mine}.
Such difficulty arises from the nested nature of ground structure problems; the inner problem is for solving the state equilibrium and the outer one is for the structural optimization using the inner solution.
That is, searching all conceivable structures is almost impossible, and hence the optimized results greatly depends on what heuristic algorithms used.

In this study, we focus on the use of quantum computers as a possible approach to tackle this mathematically hard problem.
Quantum computers have widely gained a lot of attention in recent decades due to their promise to perform quicker computations with fewer memories than those of classical computers~\cite{Harrow2009quantum}.
To deal with topology optimization problems on quantum computers, two key techniques are required.
The first is the method of solving state equilibrium equations which explain the physical phenomena we are interested in, and the second is a way for solving material distribution problems, which are intrinsically combinatorial optimization problems.
Quantum algorithms for linear systems are necessary for the first one because issues for solving the state equilibrium equation commonly reduce to problems for solving linear systems.
Several quantum algorithms for linear systems have been proposed, and some of them theoretically promise an exponential speedup over classical algorithms~\cite{Harrow2009quantum, Childs2017quantum}, whereas fault-torelant quantum computers may be required to implement them.
There also have been proposed classical-quantum hybrid algorithms for solving linear systems \cite{Bravo2019variational, Liu2021variational}, targeting the implementation of so-called noisy intermediate-scale quantum (NISQ) devices.
However, these algorithms embed the solution of linear systems into the amplitude of quantum states, which raises difficulty in efficiently extracting the solution to the classical computer.
As for the second one, on the other hand, there are several quantum algorithms for combinatorial optimization~\cite{Durr1996quantum, Chen2019optimized}, which are based on the well-known Grover's algorithm~\cite{Grover1996fast}.
There are also classical-quantum hybrid algorithms for combinatorial optimization~\cite{Vikstaal2020applying, Harwood2021formulating, Slate2021quantum}, most of which rely on the quantum approximate optimization algorithms (QAOA)~\cite{Farhi2014quantum, Zhou2020quantum, Harrigan2021quantum}.
Although bridging the difficulties of solving linear systems and combinatorial optimization is non-trivial, two crucial parts have already been identified in the literature.
We focus on the classical-quantum hybrid algorithms among these core techniques because they are predicted to execute some useful calculations using NISQ devices.
Our proposed approach is implemented using a genuine device in this study.

\begin{figure}[t]
\centering
\subfloat[][]{\includegraphics[width=0.5\linewidth]{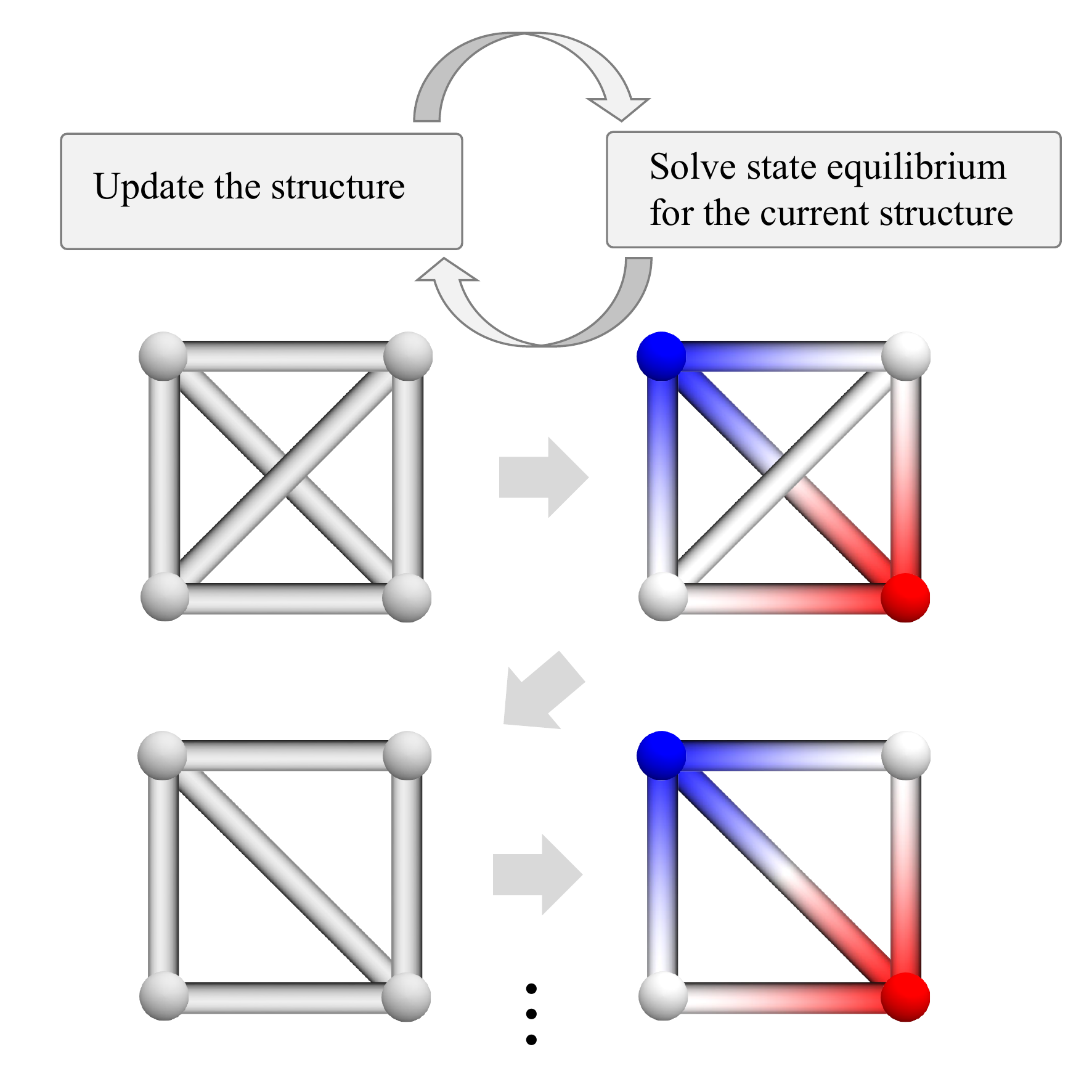} \label{subfig:classical_concept}}
\subfloat[][]{\includegraphics[width=0.5\linewidth]{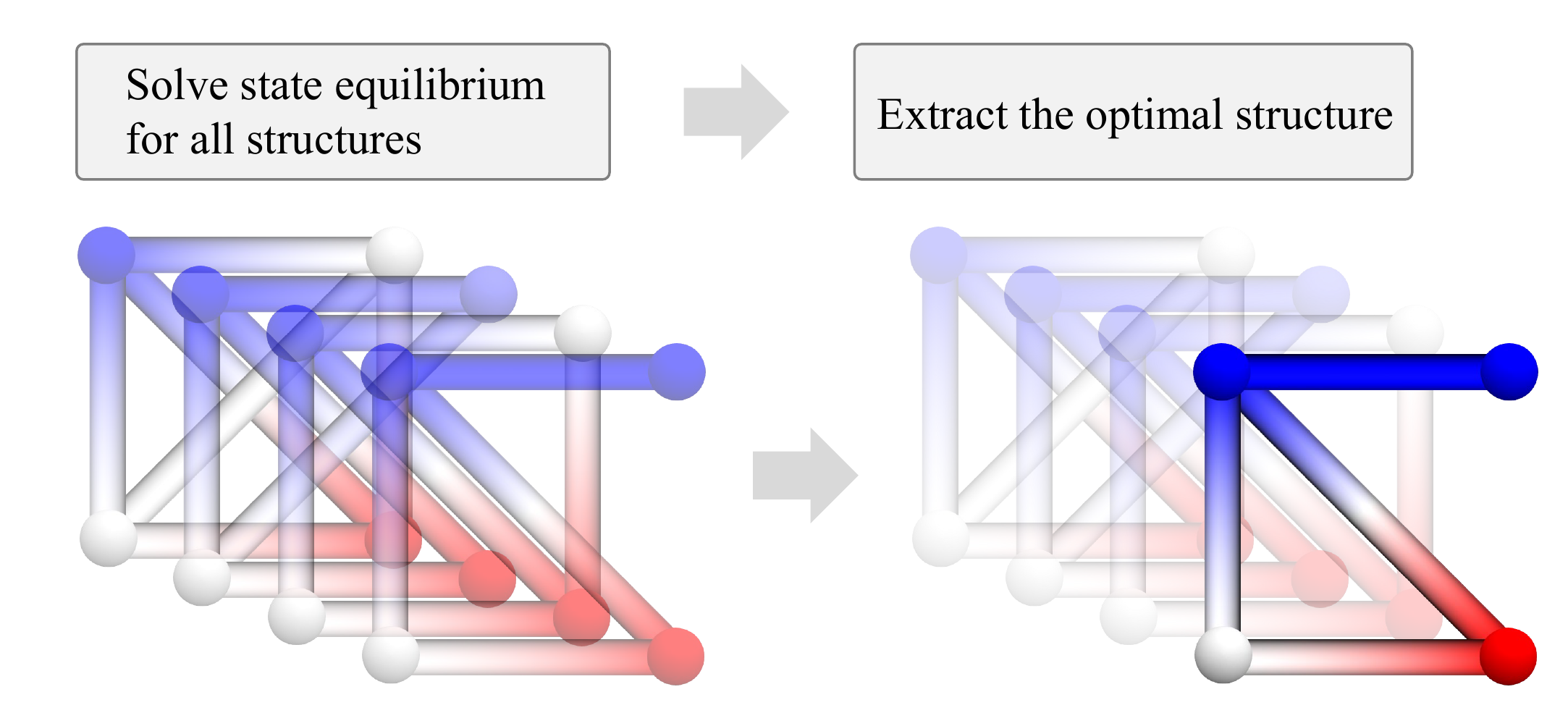} \label{subfig:quantum_concept}}
\caption{Conceptual diagrams of the ground structure methods. The color represents the equilibrium of the temperature. Red (blue) indicates hot (cold) regions. (a) Classical approach. (b) Proposed quantum approach. \label{fig:concept} }
\end{figure}

Variational quantum algorithms (VQAs)~\cite{Cerezo2021variational} are one of the most possible classical-quantum hybrid algorithms for NISQ devices.
VQAs commonly express a cost function as the expectation value of a set of observables, which is then assessed on a quantum computer utilizing a trial quantum state prepared by a parametrized quantum circuit.
To minimize the cost function, the cost function is iteratively evaluated by updating the classical parameters.
The most well-known VQAs are variational quantum eigensolvers \cite{Peruzzo2014variational, Kandala2017hardware} which were proposed to calculate the lowest eigenvalue of a system and have been extended for excited states \cite{Higgott2019variational}.
VQAs for linear systems and combinatorial optimization problems could also be found in the literature, as previously noted.
They are, insufficient for quantum computing-based topology optimization because the conventional quantum algorithms for combinatorial optimization struggle to handle the nested nature of topology optimization problems, in which the outer problem solves the structural optimization and the inner one solves the state equilibrium.
Therefore, we offer a quantum algorithm for topology optimization issues in which two VQAs are run sequentially, with the ground structure method as the aim.
The first solves the state equilibrium equation for all conceivable material configurations and prepares a quantum state in which the state variables for all possible material configurations are embedded.
Although the state variables are embedded in the amplitude of a quantum state, our focus is not to obtain the state variables themselves, but to obtain an optimal structure whose performance is calculated based on the state variables.
The second algorithm performs such optimization, that is, it prepares a parametrized quantum state describing the probability of each possible configuration and then amplifies the probability of an optimal configuration, using the quantum state obtained by the first VQA.
Figure~\ref{fig:concept} illustrates the conceptual diagrams of the conventional classical and the proposed quantum approaches for ground structure methods.
We demonstrate the effectiveness of the proposed method using a real device.

%%%%%%%%%%%%%%%%%%%%%%%%%%%%%%%%%%%%%%%%%%%%%%%%%%%%
%%%%%%%%%%%%%%%%%%%%%%%%%%%%%%%%%%%%%%%%%%%%%%%%%%%%

\section*{Results} \label{sec:results}
\subsection*{Problem definition}
This study focuses on the topology optimization problem for heat path designs~\cite{Dbouk2017review}, based on the ground structure method~\cite{Stolpe2016truss}.
Let $V:= \{ v_i \}_{i=1}^N$ and $E:= \{ e_j \}_{j=1}^m$ be the node and edge sets, respectively, where $N$ is the number of nodes and $m$ is the number of edges.
A ground structure can then be defined as the undirected graph $G=(V, E)$.
Here, we assume that each edge consists of one of two materials with different thermal conductivity values, either $\lambda$ or $\lambda'=\varepsilon \lambda$, where $\varepsilon \in (0, 1)$.
Based on this assumption, we consider the problem of assigning either material to each edge.
Let $\bs{x} = \{x_j \}_{j=1}^m \in \{0, 1\}^m$ be the structure to be optimized.
$x_j=1$ indicates that the $j$-th edge has the thermal conductivity $\lambda$, and $x_j=0$ means that $j$-th edge has the thermal conductivity $\lambda' = \varepsilon \lambda$.
Then, an optimization problem to find an optimal configuration of two kinds of edges can be formulated as follows:
\begin{alignat}{2}
	& \min_{x \in \{ 0, 1 \}^m} && \mathcal{L}(\bs{U}(x)) \label{eq:obj} \\
	& \text{subject to:} && \left(\sum_{j=1}^m K_j \left((1-\varepsilon) x_j + \varepsilon \right) \right) \bs{U} = \bs{F}, \label{eq:gov}
\end{alignat}
where $\mathcal{L}$ is the objective function to be minimized, $\bs{U}(\bs{x}) \in \mathbb{R}^N$ is the temperature vector whose $i$-th component represents the temperature of the $i$-th node at steady-state for configuration $\bs{x}$, $\bs{F} \in \mathbb{R}^N$ is the heat source vector whose $i$-th component represents a heat source at the $i$-th node, and $K_j  \in \mathbb{R}^{N \times N}$ is the element stiffness matrix of the $j$-th edge, which is determined by the material property $\lambda$, as defined in the supplementary material.
The constraint condition in Eq.~\eqref{eq:gov} represents the governing equation of steady-state heat conduction, i.e., thermal equilibrium state.
The goal is to find an optimal structure $\bs{x}^\ast$ that minimizes $\mathcal{L}(\bs{U}(\bs{x}))$.

Specifically, in the present study, we look at how to give material attributes to each edge of ground structures so that the temperature at the steady-state on a predetermined target node $v_\text{target}$ could be minimized under the existence of a heat source node $v_\text{source}$ and a heat sink node $v_\text{base}$.
In this case, the $k$-th component of the heat source vector, $F_k$, is given as
\begin{align}
    F_k = \begin{cases}
    1 & \text{if } v_k = v_\text{source} \\
    0 & \text{otherwise}
    \end{cases}. \label{eq:F_k}
\end{align}
The objective function $\mathcal{L}(\bs{U}(\bs{x}))$ is formulated as
\begin{align}
    \mathcal{L}(U(\bs{x})) = U_l(\bs{x}), \label{eq:obj_u_l}
\end{align}
where $\bs{U}_l(\bs{x})$ is the $l$-th component of the temperature vector $\bs{U}$, with $l$ being the index satisfying $v_l = v_\text{target}$.
We tackle the above problem by a two-step optimization procedure.

\subsection*{Formulation}
The use of two quantum states to represent the temperature $\bs{U}(\bs{x})$ and the structure $\bs{x}$, as well as executing separate optimization procedures to acquire these two quantum states, is a crucial feature of the present method.
In the first step, a parametrizred quantum state $\Ket{\psi (\theta)}$ is trained such that its amplitude has the information about the $i$-th component of $\bs{U}(\bs{x})$, i.e., $\braket{\bs{x}, i | \psi (\theta)} \propto \bs{U}_i (\bs{x})$ where $\Bra{\bs{x}, i}=\Bra{\bs{x}} \otimes \Bra{i}$.
Here, we summarize the result of our proposed formulation.
See Methods\nameref{sec:methods} section for detailed derivation of the optimization problems.

The problem for solving the governing equation in Eq.~\eqref{eq:gov} can be formulated as the optimization problem for encoding the state $\bs{U}(\bs{x})$ into the amplitude of $\Ket{\psi (\theta)}$, which is given as
\begin{align}
	& \min_{\bs{\theta}} \quad F_u(\bs{\theta}) = -\dfrac{\Braket{\psi(\bs{\theta})\Ket{b} \Bra{b} \psi(\bs{\theta})}}{\Braket{\psi (\bs{\theta}) | A | \psi (\bs{\theta})}}, \label{eq:opt_u}
\end{align}
where $A$ and $\Ket{b}$ are respectively defined as
\begin{equation}
	A:= \sum_{j=1}^m \dfrac{1}{2}\left((1 + \varepsilon) I^{\otimes m} - (1 - \varepsilon) Z_j \right) \otimes K_j, \label{eq:A}
\end{equation}
and
\begin{align}
	\Ket{b}:= \Ket{+}^{\otimes m} \otimes \Ket{f}, \label{eq:src}
\end{align}
where $Z_j$ represents the Pauli operator $Z$ applied to the $j$-th qubit, $\Ket{f}$ is the quantum state whose amplitude corresponds to the heat source vector $\bs{F}$, and $\Ket{+}:= \left(\Ket{0}+\Ket{1} \right) / \sqrt{2}$.
Owing to the linearity of the governing equation~\eqref{eq:gov}, we may assume that the norm of the heat source vector $\bs{F}$ is equal to $1$, without losing generality, allowing us to express the vector $\bs{F}$ as a quantum state vector $\Ket{f}$,
The heat source vector in \eqref{eq:F_k} can be represented using a quantum state as
\begin{equation}
      \Ket{f}:= \Ket{v_\text{source}},
\end{equation}
where $\Ket{v_\text{source}}$ is a vector whose component $v_\text{source}$ is $1$ corresponding to the node, while the other components are $0$.
The optimized parameter $\bs{\theta}^\ast$ is now fixed in the following step.

In the second step, we present another parametrized quantum state $\Ket{\phi (\eta)}$ whose bit string outcome is measured by the computational basis that matches with the structure $\bs{x}$.
Let $O$ denote a Hermitian operator defined as
\begin{equation}
      O:= \Ket{v_\text{target}} \Bra{v_\text{target}}, \label{eq:opt_s_o}
\end{equation}
where $\Ket{v_\text{target}}$ represents a vector in which the component corresponding to the node $v_\text{target}$ is $1$ and the other components are $0$.
Then we redefine the objective function in Eq.~\eqref{eq:obj_u_l} as follows:
\begin{align}
      \mathcal{L}(\bs{U}(\bs{x})) = \bs{U}^\dagger O \bs{U}, \label{eq:obj_specific}
\end{align}
where $\dagger$ represents the Hermitian transpose.
Now we consider the weighted sum of the objective function values for all possible structures and optimize the weighting coefficients to minimize the weighted sum value, resulting in only the weighting coefficient corresponding to the optimal solution with the minimum objective function value being $1$ and anything else $0$.
For all feasible structures, the weighted sum of the objective function values can be stated as
\begin{align}
      & \sum_{\bs{x} \in \{0,1\}^m} P(\bs{x}) \bs{U}(\bs{x})^\dagger O \bs{U}(\bs{x}) \propto \sum_{\bs{x} \in \{0,1\}^m} P(\bs{x}) \Braket{\psi (\bs{\theta}^\ast) | \left(\Ket{x} \Bra{x} \otimes O \right) | \psi (\bs{\theta}^\ast) },
\end{align}
where $P(\bs{x})$ represents the probability of each possible structure.
Hence, the objective function describes the expectation value of the target node temperature over all the ground structures. 
It is obvious that the optimal $P(\bs{x})$ corresponds to the optimal solution $\bs{x}^\ast$ of the original problem.
Parametrizing $P(\bs{x})$ by parameters $\bs{\eta}$ through a parametrized quantum state $\Ket{\phi (\eta)}$, i.e., $P_\eta (\bs{x}):= \left| \Braket{x | \phi (\bs{\eta})} \right|^2$, we can formulate the optimization problem for obtaining the optimal structure as follows:
\begin{align}
      & \min_{\bs{\eta}} \quad F_s(\bs{\eta}) = \sum_{\bs{x} \in \{0,1\}^m} P_\eta (\bs{x}) \Braket{\psi (\bs{\theta}^\ast) | \left(\Ket{x} \Bra{x} \otimes O \right) | \psi (\bs{\theta}^\ast) }. \label{eq:opt_s}
\end{align}
The quantum state $\Ket{\phi (\eta^\ast)}$ is measured once the optimized parameters $\bs{\eta}^\ast$ are obtained.
When all the preceding steps are completed successfully, the likelihood of identifying optimum structures $\bs{x}^\ast$ will become $1$.
See Methods\nameref{sec:methods} section for detailed derivation of Eq.~\eqref{eq:opt_s}.

%%%%%%%%%%%%%%%%%%%%%%%%%%%%%%%%%%%%%%%%%%%%%%%%%%%%
\begin{figure}[t]
\centering
\subfloat[][]{\includegraphics[width=0.5\linewidth]{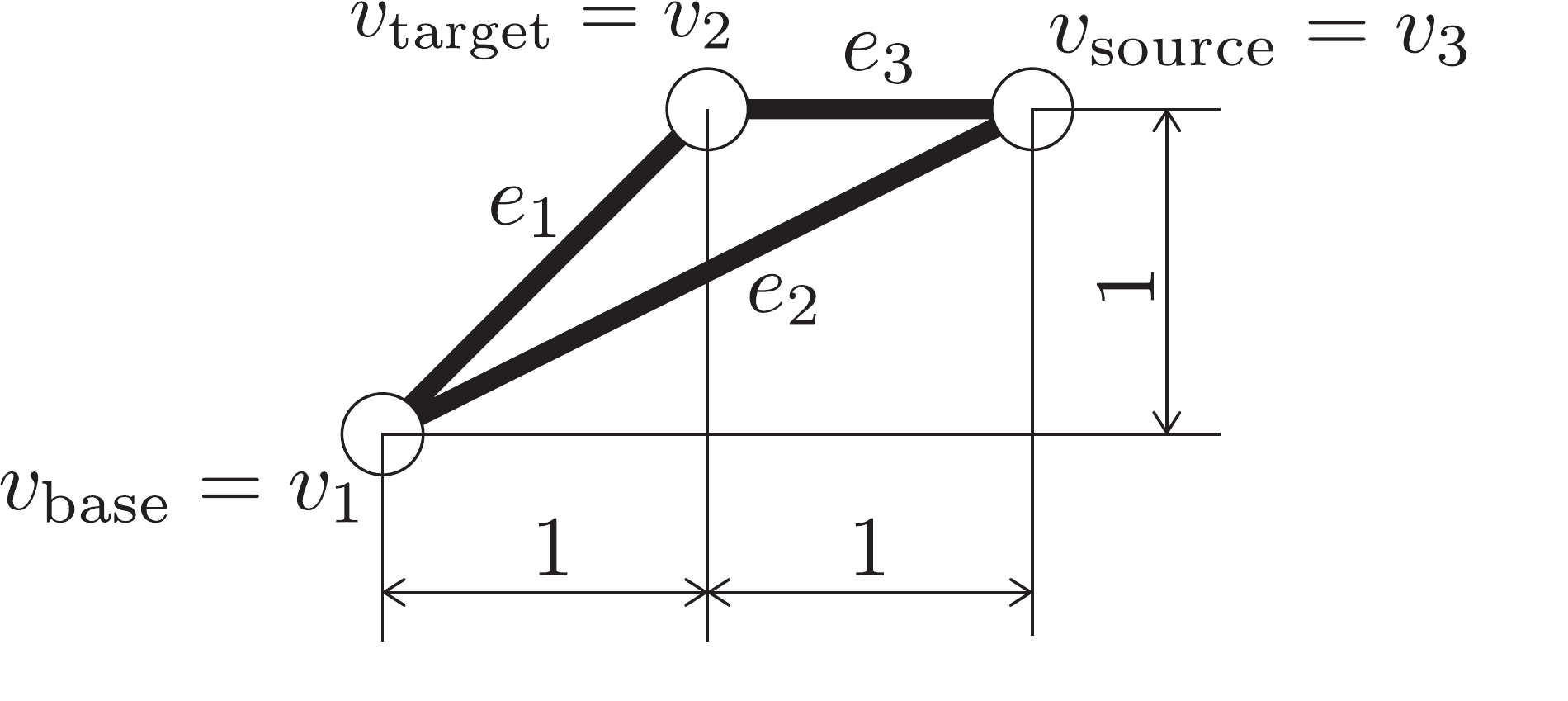} \label{subfig:setting}}
\subfloat[][]{\includegraphics[width=0.3\linewidth]{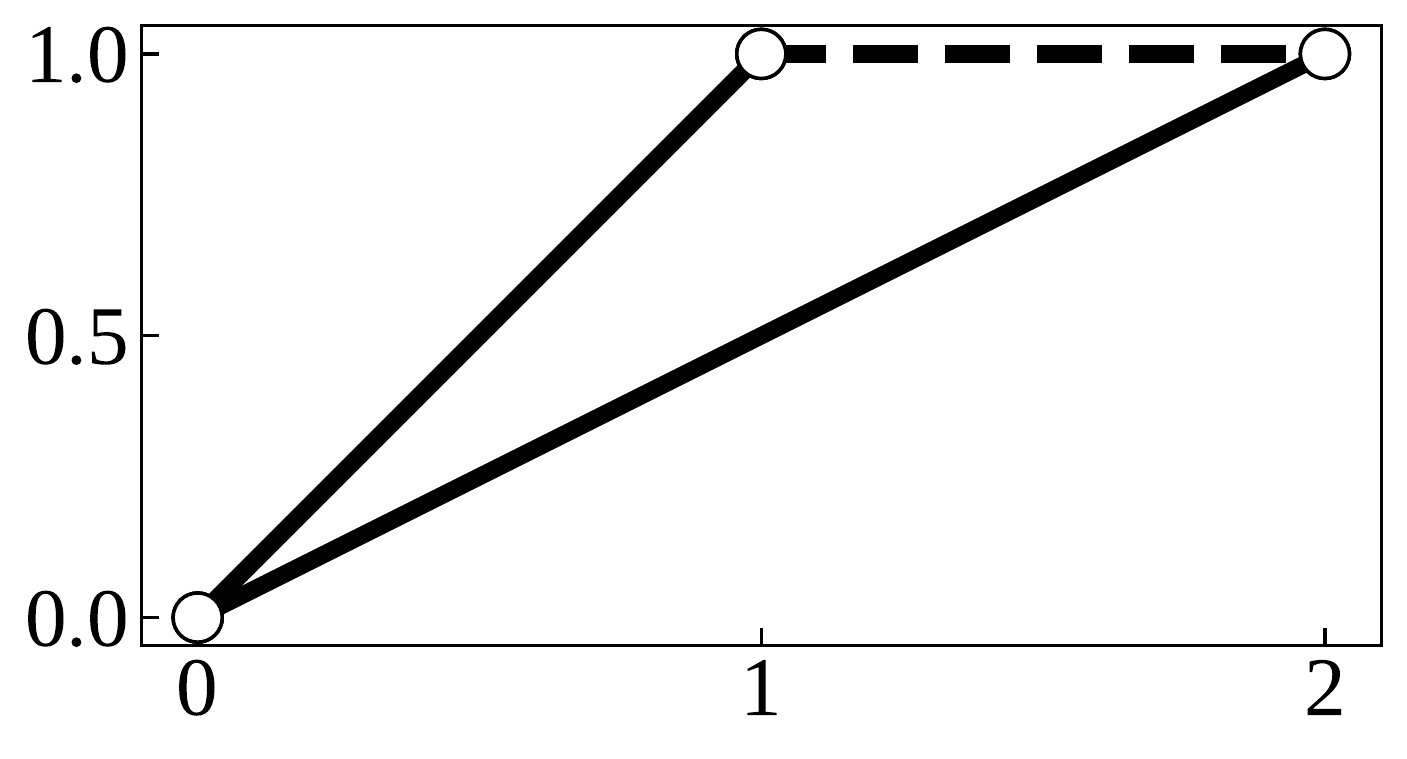} \label{subfig:optstr}}
\caption{ The problem of three edges-design. (a) The settings of the problem. (b) The optimal structure which is represented by the bit string 110. The solid lines represent materials with higher thermal conductivity, i.e., $x_j=1$, while the dashed line represents the material with smaller thermal conductivity, i.e., $x_j=0$. \label{fig:3edges_prob} }
\end{figure}
%%%%%%%%%%%%%%%%%%%%%%%%%%%%%%%%%%%%%%%%%%%%%%%%%%%%

\subsection*{Three-edge-design-problem}
%%%%%%%%%%%%%%%%%%%%%%%%%%%%%%%%%%%%%%%%%%%%%%%%%%%%
\begin{figure}[!h]
	\centering
	\subfloat[][]{\includegraphics[width=0.3\linewidth]{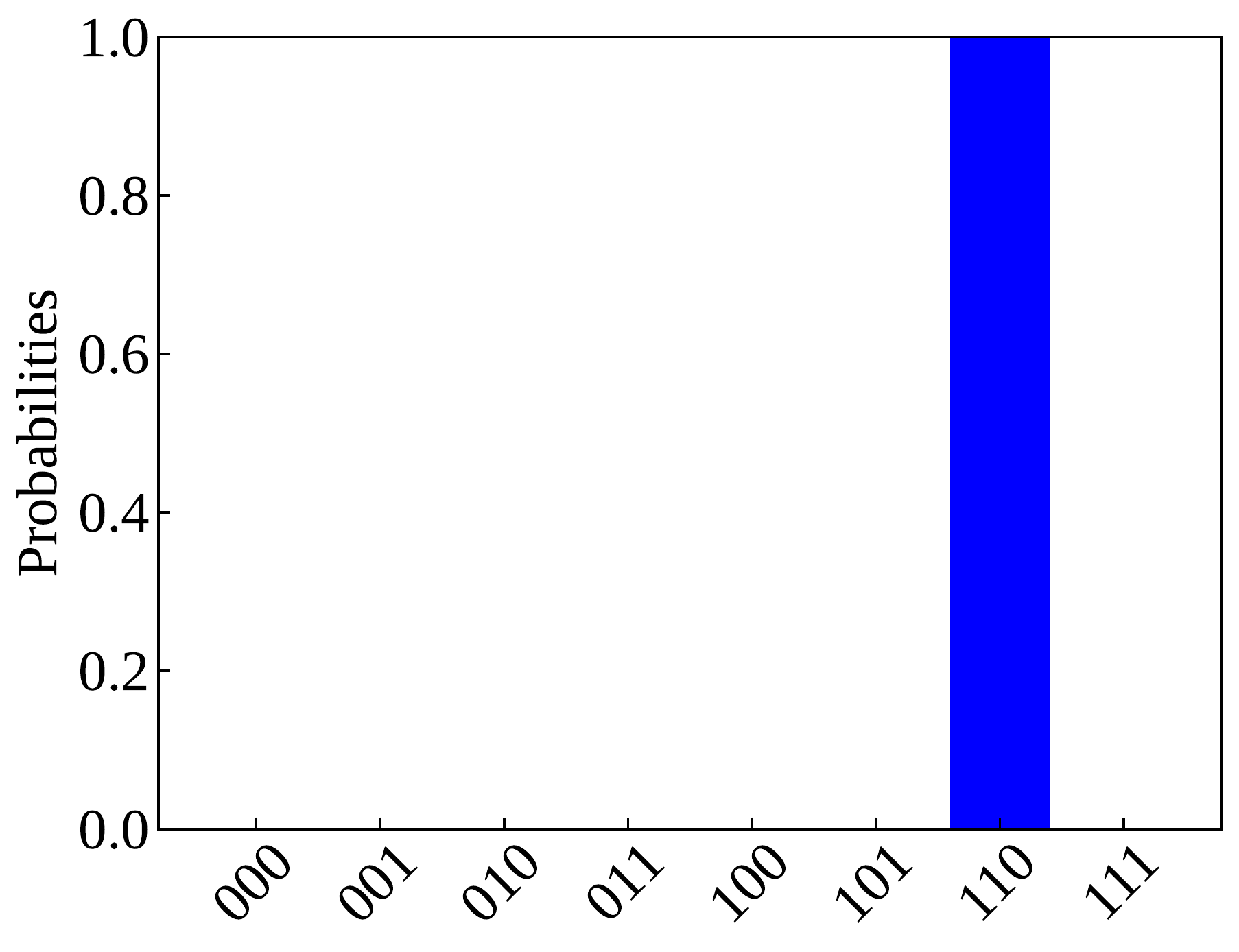} \label{subfig:3edges_probability}} \quad
	\subfloat[][]{\includegraphics[width=0.3\linewidth]{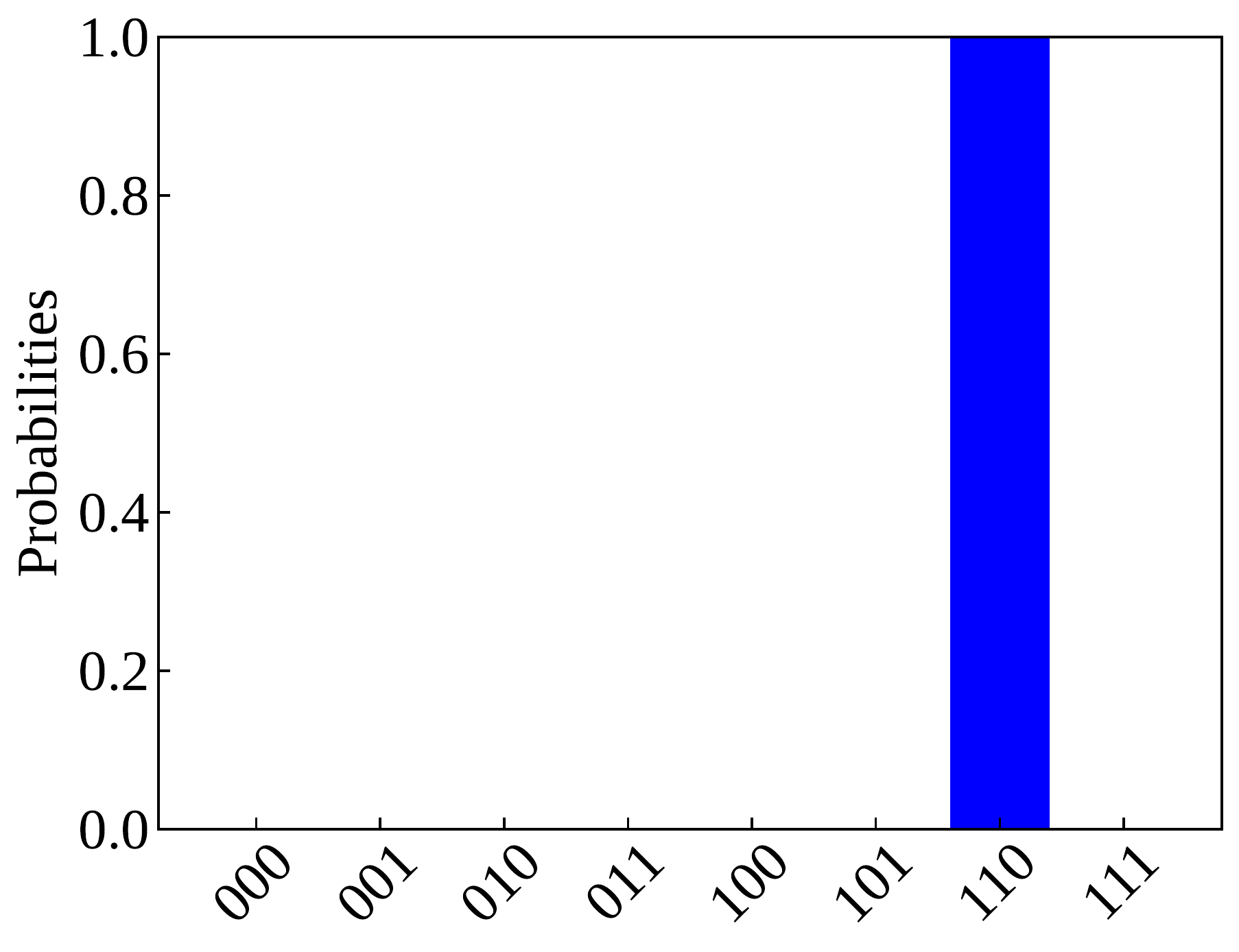} \label{subfig:3edges_shots_probability}} \quad
	\subfloat[][]{\includegraphics[width=0.3\linewidth]{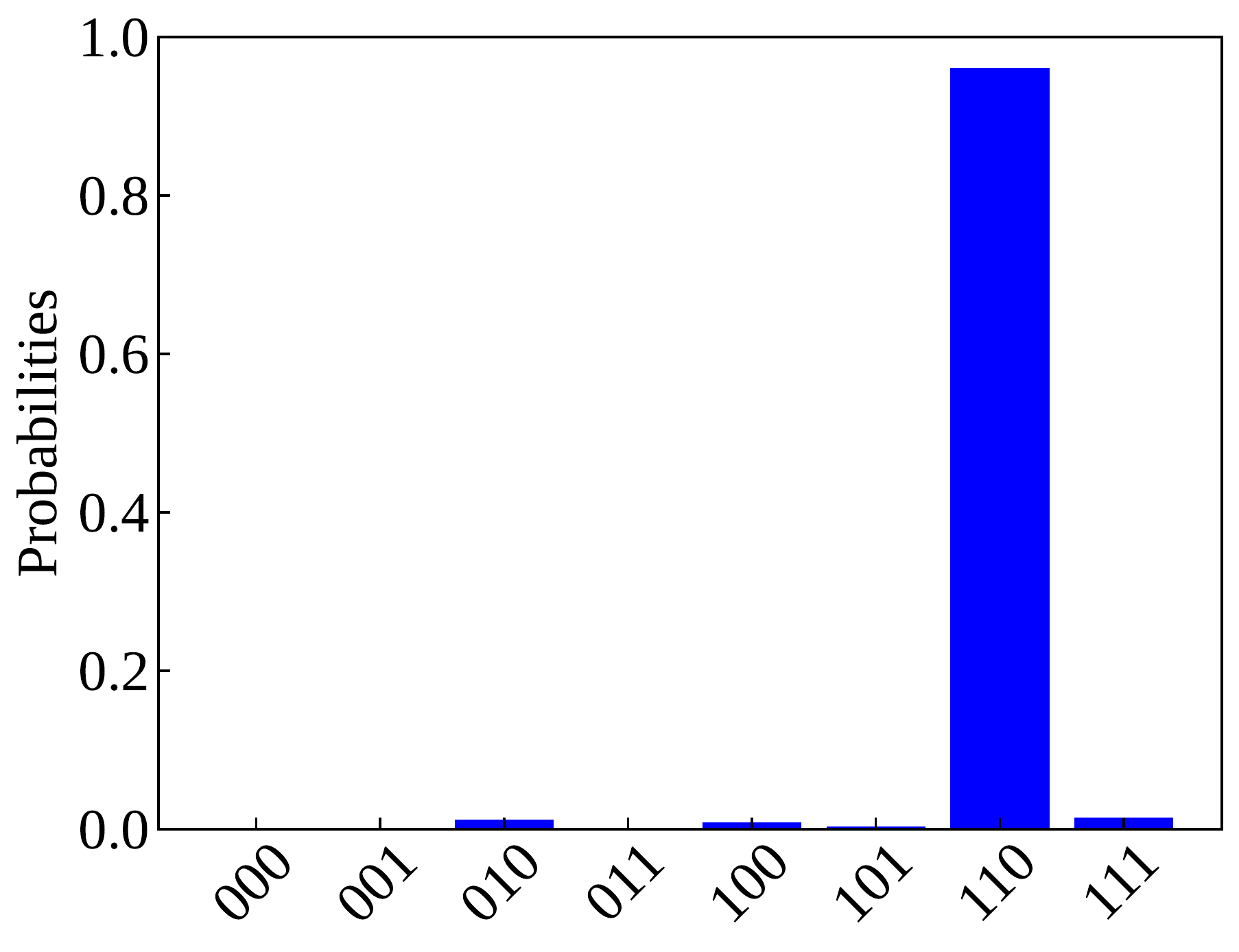} \label{subfig:3edges_kawasaki_probability}} \\
	\subfloat[][]{\includegraphics[width=0.3\linewidth]{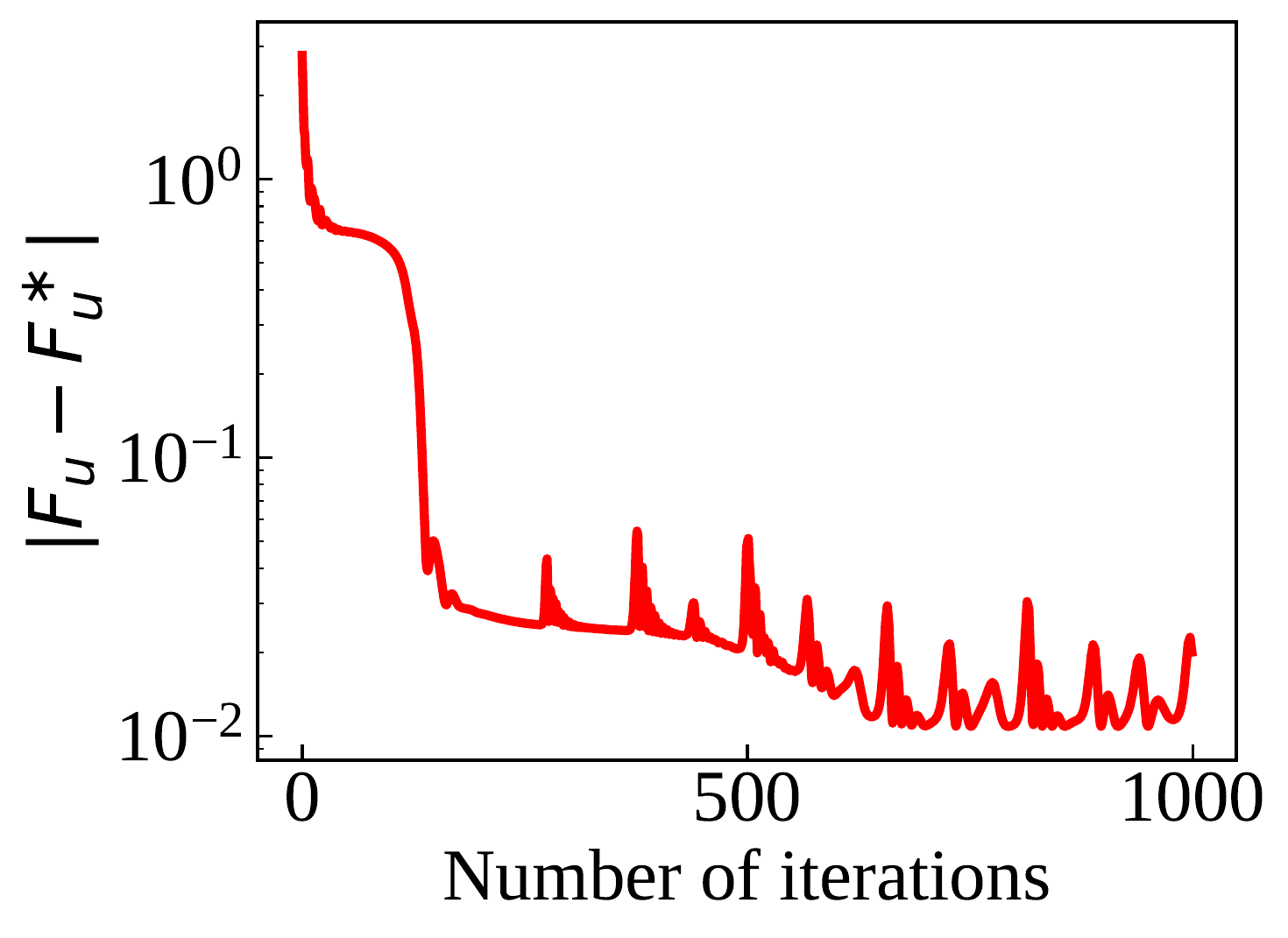} \label{subfig:3edges_history_state}} \quad
	\subfloat[][]{\includegraphics[width=0.3\linewidth]{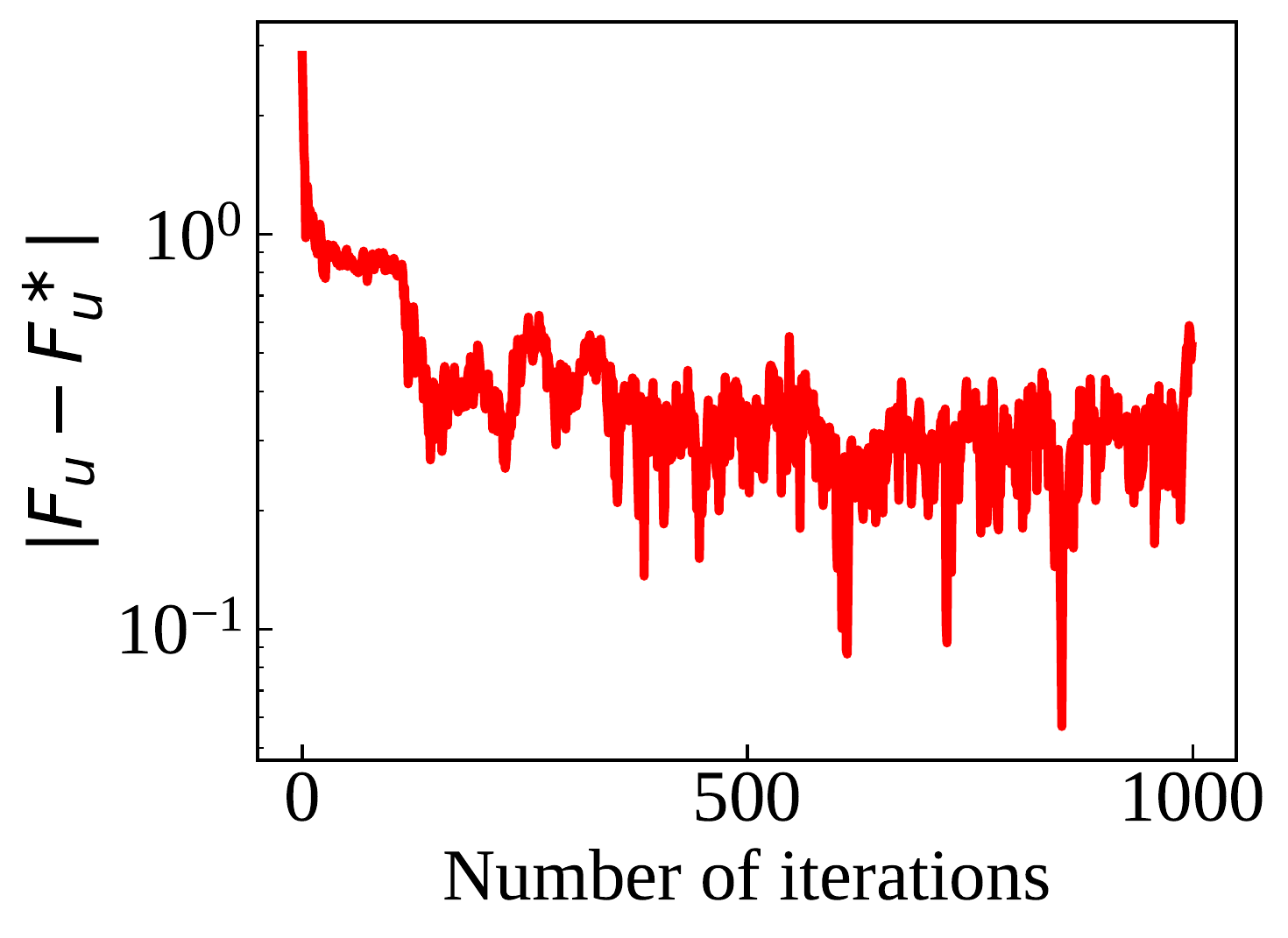} \label{subfig:3edges_shots_history_state}} \quad
	\subfloat[][]{\includegraphics[width=0.3\linewidth]{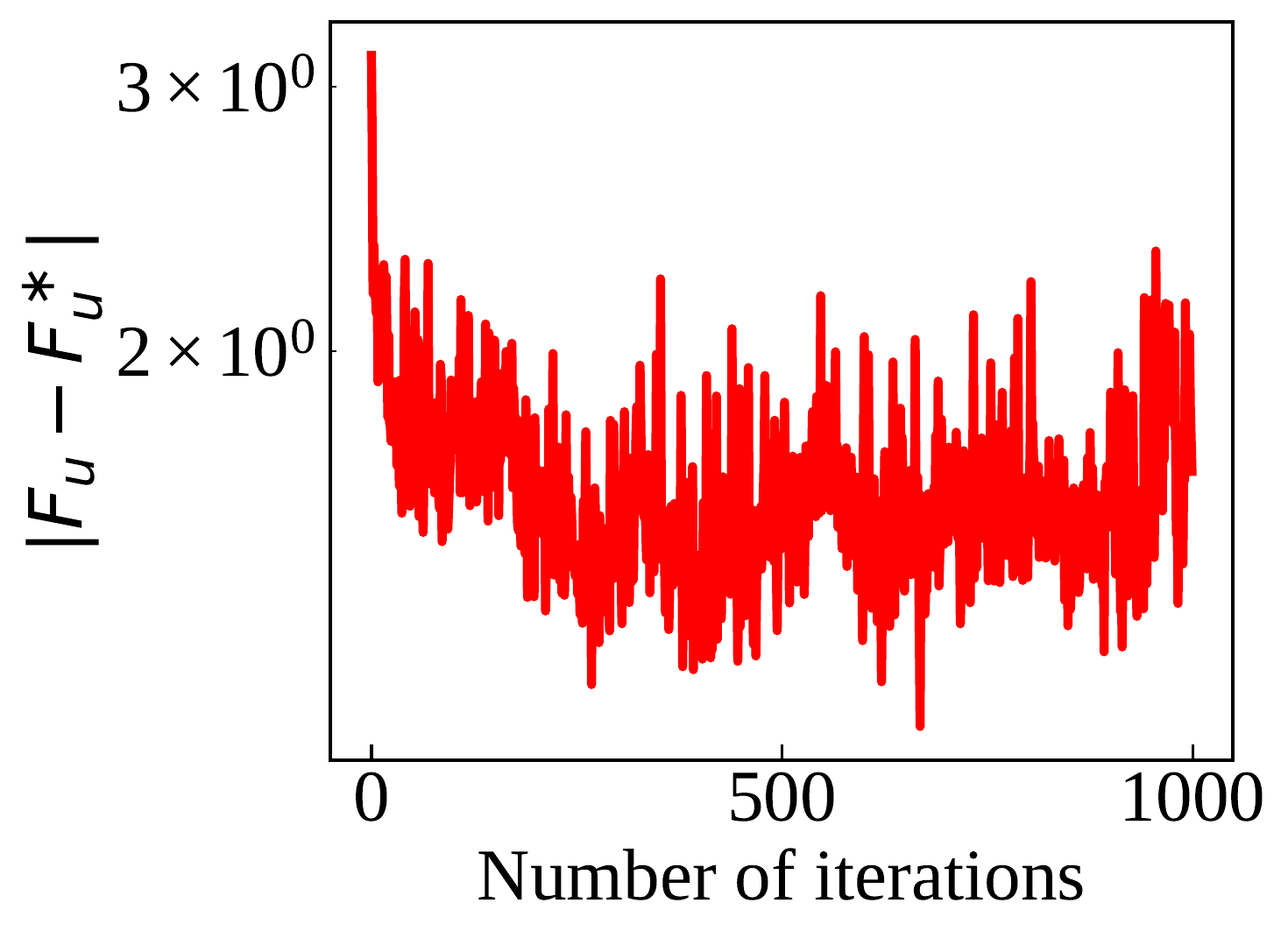} \label{subfig:3edges_kawasaki_history_state}} \\	
	\subfloat[][]{\includegraphics[width=0.3\linewidth]{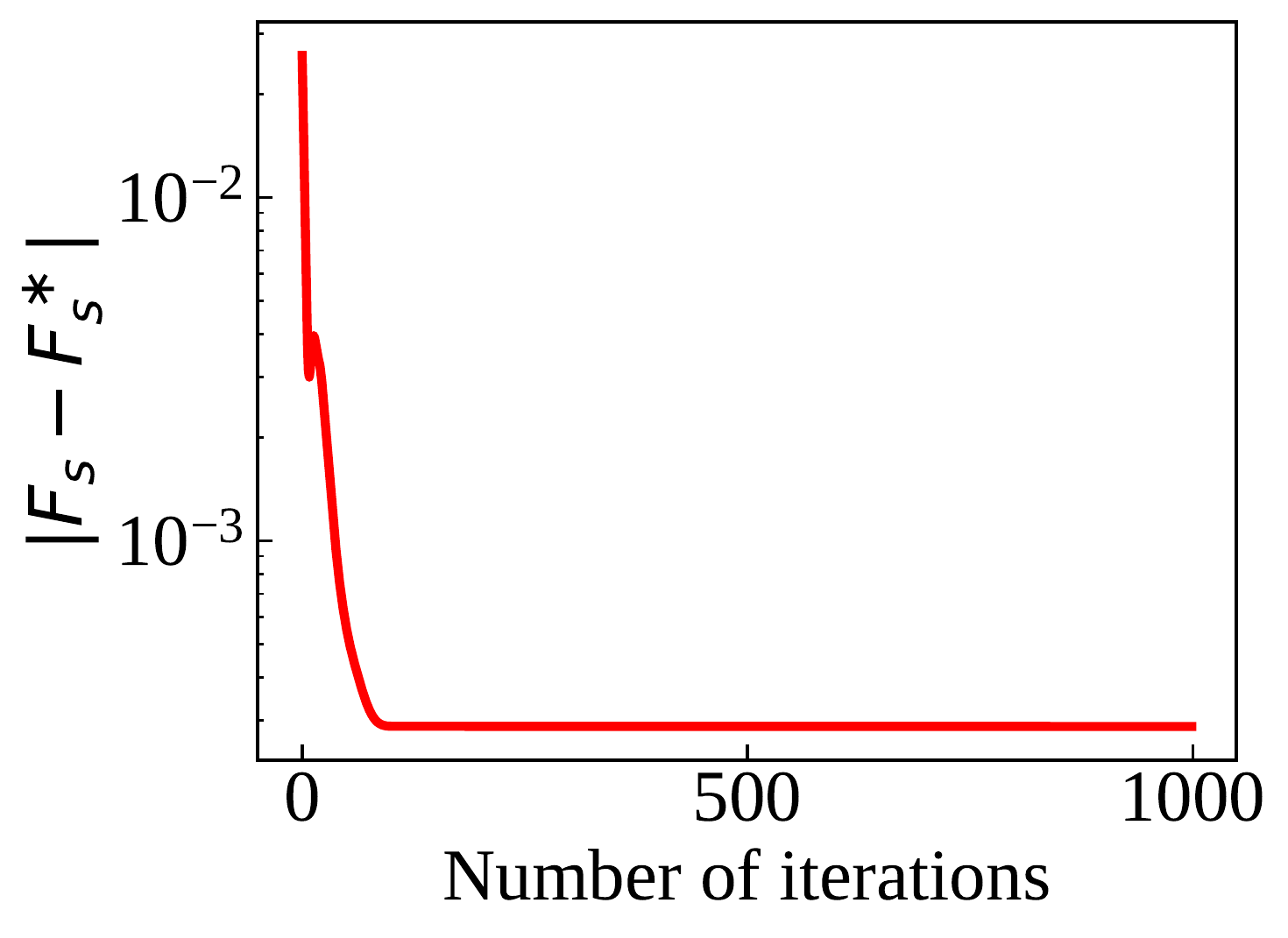} \label{subfig:3edges_history_struct}} \quad	
	\subfloat[][]{\includegraphics[width=0.3\linewidth]{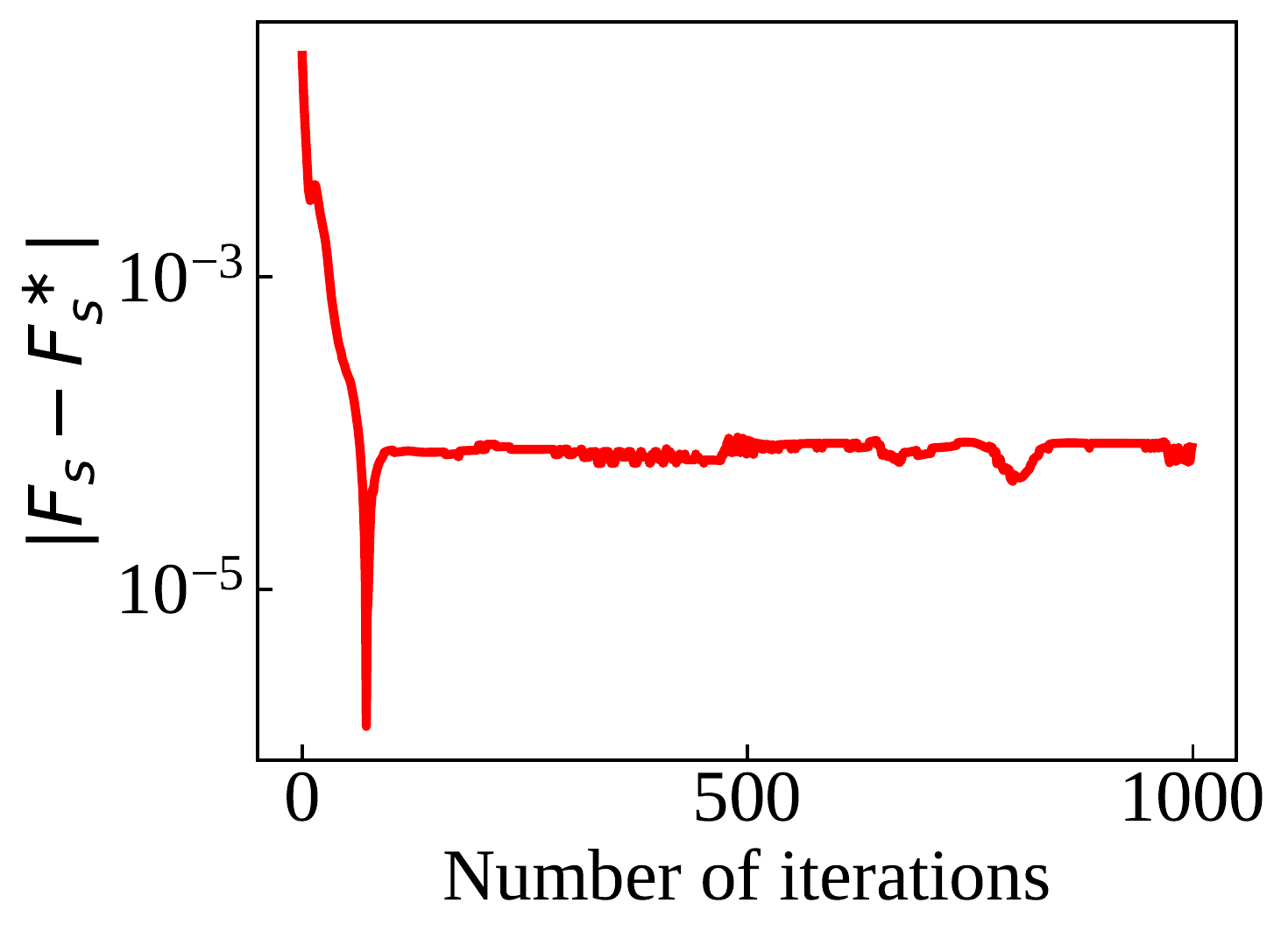} \label{subfig:3edges_shots_history_struct}} \quad
	\subfloat[][]{\includegraphics[width=0.3\linewidth]{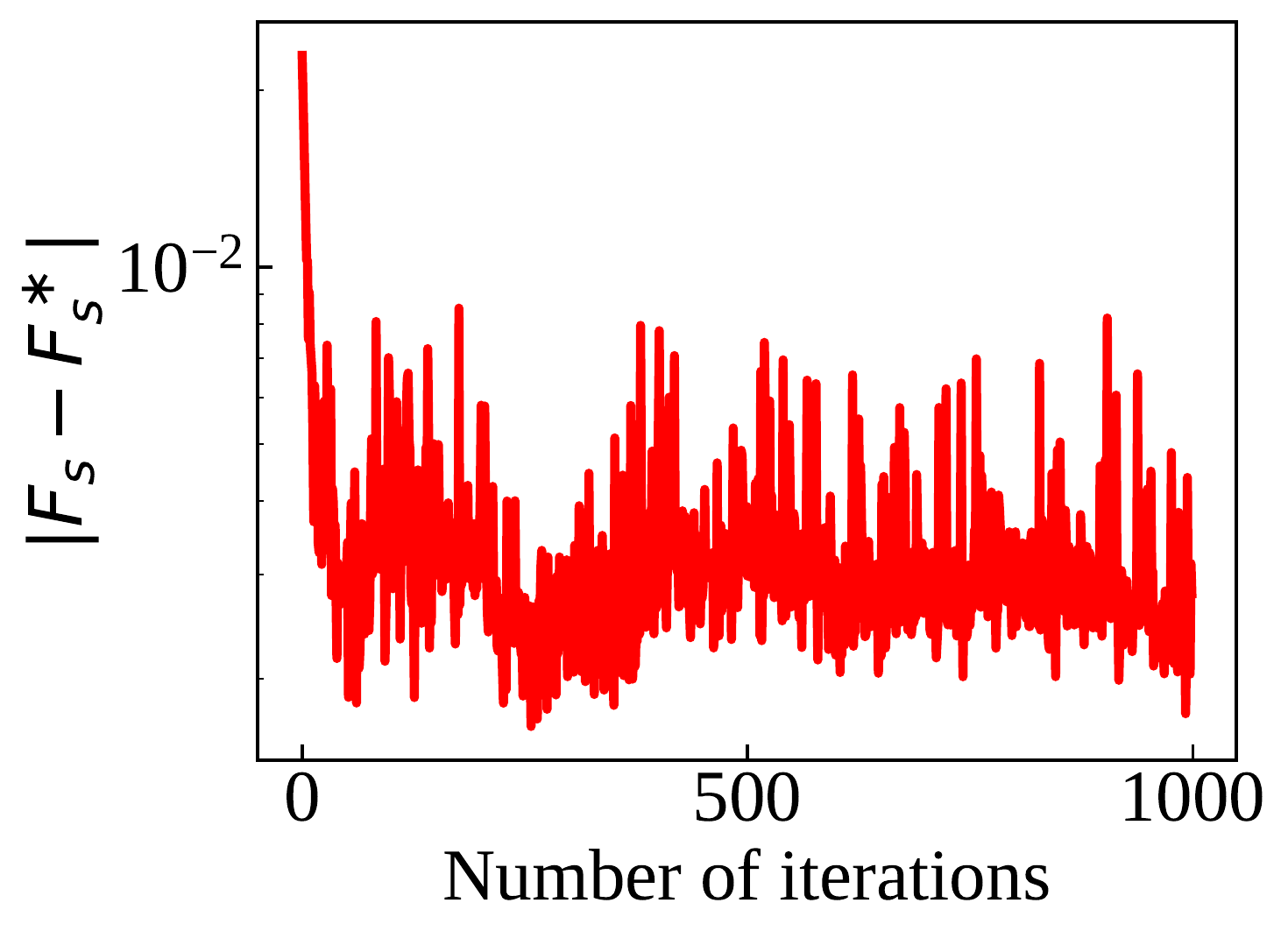} \label{subfig:3edges_kawasaki_history_struct}}
	\caption[]{Result of the three edges-design problems. Probability distribution of the measurement outcomes of $\Ket{\phi (\bar{\eta})}$, where $\bar{\eta}$ is the optimized parameters by (a) the \textit{statevector simulator}, (b) the \textit{QASM simulator}, and (c) the real device (\textit{ibm\_kawasaki}). Bit strings in the horizontal axis represents the structures $\{x_j \}_{j=1}^3$. 
	History of $|F_u - F_u^\ast|$ obtained by using (d) the \textit{statevector simulator}, (e) the \textit{QASM simulator}, and (f) the real device (\textit{ibm\_kawasaki}), where $F_u^\ast$ is the minimum value of $F_u$ in Eq.~\eqref{eq:obj_u}. 
	History of $|F_s - F_s^\ast|$ obtained by using (g) the \textit{statevector simulator}, (h) the \textit{QASM simulator}, and (i) the real device (\textit{ibm\_kawasaki}), where $F_s^\ast$ is the minimum value of $F_s$ in Eq.~\eqref{eq:obj_s}. \label{fig:3edges_result}}
\end{figure}
%%%%%%%%%%%%%%%%%%%%%%%%%%%%%%%%%%%%%%%%%%%%%%%%%%%%
First, we look at a design challenge that involves giving material properties to three edges, as shown in Fig.~\ref{fig:3edges_prob} \subref{subfig:setting}.
The temperature on the node $v_\text{target}$ is an objective function to be minimized under the condition that the temperature on the node $v_\text{base}$ is set to $0$ and the heat source is applied on the node $v_\text{source}$.
The optimal structure which is obtained by classical brute-force search is shown in Fig.~\ref{fig:3edges_prob} \subref{subfig:optstr}.
We utilized the alternating layered ansatz~\cite{Cerezo2021cost} whose number of layers was set to $2$ for creating the state $\Ket{\psi (\theta)}$ and $1$ for the state $\Ket{\phi (\eta)}$.
In the supplementary material, there is another example where material properties are assigned to five edges.

The thermal conductivity of a material was set to $\lambda=1$ in all the subsequent experiments, and the ratio of the thermal conductivity of the other material to the material was set to $\varepsilon=0.1$.
Qiskit~\cite{Qiskit}Ver. 0.32.1, an open-source framework for working with quantum computers, was used to build the proposed method.
The adaptive moment estimation (ADAM)~\cite{Kingma2014adam} was utilized as a parameterized quantum circuits optimizer.
The learning rate was set to $0.1$, the hyperparameters $\beta_1$ and $\beta_2$ were respectively set to $0.9$ and $0.999$, and the number of iterations was set to $1,000$ for all experiments.
Except for the case using the \textit{statevector simulator} in Qiskit~\cite{Qiskit}, the number of shots for each quantum circuit was set to $32,000$.

\subsubsection*{Statevector simulation}
%%%%%%%%%%%%%%%%%%%%%%%%%%%%%%%%%%%%%%%%%%%%%%%%%%%%
% \begin{figure}[t]
% \centering
% \subfloat[][]{\includegraphics[width=0.4\linewidth]{_plot_adam_3edges.pdf} \label{subfig:3edges_probability}} \quad
% \subfloat[][]{\includegraphics[width=0.3\linewidth]{_plot_110_adam_3edges.pdf} \label{subfig:3edges_optstr}} \\
% \subfloat[][]{\includegraphics[width=0.4\linewidth]{_plot_history_state_obj_3edges.pdf} \label{subfig:3edges_history_state}} \quad
% \subfloat[][]{\includegraphics[width=0.4\linewidth]{_plot_history_struct_obj_3edges.pdf} \label{subfig:3edges_history_struct}}
% \caption[]{Result of the three-edges-design problem by \textit{statevector simulator}. \subref{subfig:3edges_probability} Probability distribution of the measurement outcomes of $\Ket{\phi (\bar{\eta})}$ where $\bar{\eta}$ is the optimized parameters. The bit strings in the horizontal axis represents the structures $\{x_j \}_{j=1}^3$. \subref{subfig:3edges_optstr} The structure corresponding to the bit string 110. The bold edges represents the material with bigger thermal conductivity, i.e., $x_j=1$, while the thin edge represents the material with smaller thermal conductivity, i.e., $x_j=0$. This structure coincides to the exact optimal structure obtained by the brute-force search. \subref{subfig:3edges_history_state} The history of $F_u$ in Eq.~\eqref{eq:obj_u}. \subref{subfig:3edges_history_struct} The history of $F_s$ in Eq.~\eqref{eq:obj_s}. \label{fig:3edges_result}}
% \end{figure}
%%%%%%%%%%%%%%%%%%%%%%%%%%%%%%%%%%%%%%%%%%%%%%%%%%%%
The \textit{statevector simulator} backend was used for the numerical experiments in this section.
The result of the three edges-design problem obtained using the \textit{statevector simulator} is shown in Fig.~\ref{fig:3edges_result}.
As shown in Fig.~\ref{fig:3edges_result}\subref{subfig:3edges_probability}, the probability concentrates on a structure, which agrees with the exact optimal structure illustrated in Fig.~\ref{fig:3edges_prob} \subref{subfig:optstr}.
Figures~\ref{fig:3edges_result}\subref{subfig:3edges_history_state} and \subref{subfig:3edges_history_struct} show the histories of two optimization problems, one of which is for calculating the response of the system, and the other of which is for amplifying the probability of the optimal structure.
The objective function value rapidly declines in this image, which is accompanied by oscillations.

\subsubsection*{QASM simulation}
%%%%%%%%%%%%%%%%%%%%%%%%%%%%%%%%%%%%%%%%%%%%%%%%%%%%
% \begin{figure}[t]
% \centering
% \subfloat[][]{\includegraphics[width=0.4\linewidth]{_plot_adam_3edges_shots.pdf} \label{subfig:3edges_shots_probability}} \\
% \subfloat[][]{\includegraphics[width=0.4\linewidth]{_plot_history_state_obj_3edges_shots.pdf} \label{subfig:3edges_shots_history_state}} \quad
% \subfloat[][]{\includegraphics[width=0.4\linewidth]{_plot_history_struct_obj_3edges_shots.pdf} \label{subfig:3edges_shots_history_struct}}
% \caption[]{Result of the three-edges-design problem by \textit{QASM simulator}. \subref{subfig:3edges_shots_probability} Probability distribution of the measurement outcomes of $\Ket{\phi (\bar{\eta})}$ where $\bar{\eta}$ is the optimized parameters for the three-edges-design problem solved by the \textit{QASM simulator}. The bit strings in the horizontal axis represents the structures $\{x_j \}_{j=1}^3$. \subref{subfig:3edges_shots_history_state} The history of $F_u$ in Eq.~\eqref{eq:obj_u}. \subref{subfig:3edges_shots_history_struct} The history of $F_s$ in Eq.~\eqref{eq:obj_s}. \label{fig:3edges_shots_result}}
% \end{figure}
%%%%%%%%%%%%%%%%%%%%%%%%%%%%%%%%%%%%%%%%%%%%%%%%%%%%
The \textit{QASM simulator} backend, which can simulate sampling by measurements without any noise, was used for the numerical experiments in this section.
The probability distributions for each faesible structure at the end of the optimization process are shown in Fig.~\ref{fig:3edges_result}\subref{subfig:3edges_shots_probability}.
This figure shows that the probability concentrates on the bit string $110$, which coincides with the actual optimal structure, obtained by the brute-force search, as shown in Fig.~\ref{fig:3edges_prob} \subref{subfig:optstr}.
Figures~\ref{fig:3edges_result}\subref{subfig:3edges_shots_history_state} and \subref{subfig:3edges_shots_history_struct} demonstrate the two optimization histories, one of which is for acquiring the response of the system, and the other for increasing the probability of the optimal structure.
The objective function value of $F_u$ decreased with oscillations, when compared with the case of utilizing the \textit{statevector simulator}. 
It can be observed that this is due to the statistical error in calculating the objective and its gradient by the finite number of sampling.

\subsubsection*{Real device experiment}
%%%%%%%%%%%%%%%%%%%%%%%%%%%%%%%%%%%%%%%%%%%%%%%%%%%%
% \begin{figure}[t]
% \centering
% \subfloat[][]{\includegraphics[width=0.4\linewidth]{_plot_adam_3edges_kawasaki.pdf} \label{subfig:3edges_kawasaki_probability}} \\
% \subfloat[][]{\includegraphics[width=0.4\linewidth]{_plot_history_state_obj_3edges_kawasaki.pdf} \label{subfig:3edges_kawasaki_history_state}} \quad
% \subfloat[][]{\includegraphics[width=0.4\linewidth]{_plot_history_struct_obj_3edges_kawasaki.pdf} \label{subfig:3edges_kawasaki_history_struct}}
% \caption[]{Result of the three-edges-design problem by \textit{ibm\_kawasaki}. \subref{subfig:3edges_kawasaki_probability} Probability distribution of the measurement outcomes of $\Ket{\phi (\bar{\eta})}$ where $\bar{\eta}$ is the optimized parameters for the three-edges-design problem solved by the \textit{ibm\_kawasaki}. The bit strings in the horizontal axis represents the structures $\{x_j \}_{j=1}^3$. \subref{subfig:3edges_kawasaki_history_state} The history of $F_u$ in Eq.~\eqref{eq:obj_u}. \subref{subfig:3edges_kawasaki_history_struct} The history of $F_s$ in Eq.~\eqref{eq:obj_s}. \label{fig:3edges_kawasaki_result}}
% \end{figure}
%%%%%%%%%%%%%%%%%%%%%%%%%%%%%%%%%%%%%%%%%%%%%%%%%%%%
The \textit{ibm\_kawasaki} backend in IBM Quantum~\cite{IBMQ} was used for the experiments in this section.
Figure~\ref{fig:3edges_result}\subref{subfig:3edges_kawasaki_probability} illustrates the probability distributions for each possible structure sampled after the optimization process halts.
This figure shows that the probability concentrates on the structure $110$, which coincides with the actual optimal structure, obtained by the brute-force search, shown in Fig.~\ref{fig:3edges_prob} \subref{subfig:optstr}.
The probability did not reach $1.0$ different comparison to the case of utilizing the simulator, which appears to be due to the noise.
The probability of the structure $110$ was $0.956$.
Figures~\ref{fig:3edges_result}\subref{subfig:3edges_kawasaki_history_state} and \subref{subfig:3edges_kawasaki_history_struct} show the two optimization histories, one of which is for solving the state equilibrium, and the other of which is for solving the structural optimization.
Both the objective function values dropped with significant oscillations as compared to the case of utilizing the simulator, which is attributable to the noise in addition to the statistical error in calculating the objective and its gradient.
Nonetheless, we believe that the proposed method is adequate for NISQ devices because it can get an actual optimal structure in this small-scale problem of utilizing $4$ qubits.

\section*{Discussion} \label{sec:discussion}
The proposed method requires $m + \lceil \log_2 N \rceil$ qubits to represent the temperature vector for all possible structures, while classical approaches require $\mathcal{O}(N)$ bits to store the temperature vector for a certain structure.
When we assume that the ground structures are given as $k$-regular graph, the number of edges is given as $m=kN/2$.
Therefore, the number of qubits required in the proposed method is $kN/2 + \lceil \log_2 N \rceil$, which is roughly the same order as that in the classical approaches with respect to $N$ when $N$ is large.
Because the proposed method deals with all structures simultaneously, whereas the classical approaches usually deal with a certain structure at once, this scaling is an advantage of the proposed approach in terms of the resources required.
When we treat all structures simultaneously in the classical approaches, we require $\mathcal{O}(N) 2^m$ bits to store the temperature vector for all structures at once.
Actually, in the classical approaches, it is not required to store the temperature vector for all structures at once, while $2^m$ calculations are required for the brute-force approach.

The results show that the proposed method is capable of obtaining the exact optimal structure for the three edges-design problem.
The key property of the proposed method is that no amplitude estimation with respect to $\Ket{\Psi (\theta)}$ is required.
Although existing quantum linear solvers including the HHL algorithm~\cite{Harrow2009quantum} may efficiently solve linear systems, the solutions are encoded in the amplitude of quantum states.
To decode the solution to the classical computer, amplitude estimation~\cite{Brassard2002quantum, Tanaka2021amplitude} or quantum state tomography~\cite{Xin2019local, Liu2020variational} is necessary.
The proposed method, in contrast, leverages the quantum state encoding the linear system solutions as the second step optimization module, and the solution we want to extract to the classical computer is the bit string representing the optimal structure.
Hence, amplitude estimation and quantum state tomography are not required.
This would be a noteworthy feature of the proposed method.
As a result of the experimental findings, it was discovered that the effect of the noise and the statistical error created severe oscillations, potentially raising the issue of scalability.
In the future, we will perform a detailed analysis of these factors' impact on scalability.

We believe that our proposed method can also be applied to other research fields where mixed integer programming problems exist, given the mathematical structure of the topology optimization problems based on the ground structures.

\section*{Methods} \label{sec:methods}
\subsection*{Formulation for quantum computing}
Here, we formulate two optimization problems to solve the original optimization problem in Eqs.~\eqref{eq:obj} and \eqref{eq:gov} based on quantum computing.
The one is for solving the governing equation~\eqref{eq:gov}, and is introduced in the following section.
The other is for the original optimization problem in Eqs.~\eqref{eq:obj} and \eqref{eq:gov}, which is formulated below.

\subsubsection*{Solving the linear system $\bs{K}\bs{U} = \bs{F}$ by variational quantum algorithm}
First, we derive an optimization problem for using quantum computing to solve the governing equation~\eqref{eq:gov}.
We assume that $N=2^n$, which means that the governing equation with the dimension of $N$ can be handled in a $n$-qubits quantum system.
In the supplemental information, the case where this assumption does not hold is examined.

Let $u_i^{(\bs{x})}$ be defined as the $i$-th node temperature of the structure $\bs{x}$.
The vector $\Ket{\Psi}$ is then defined as follows:
\begin{align}
      \Ket{\Psi}:= \dfrac{1}{2^{m/2}} \sum_{\bs{x} \in \{0, 1\}^m} \sum_{i=1}^{N} u_i^{(\bs{x})} \Ket{\bs{x}} \otimes \Ket{i},
\end{align}
which is the solution of the following linear system:
\begin{equation}
      A \Ket{\Psi} = \Ket{b}, \label{eq:gov_q}
\end{equation}
where $A$ and $\Ket{b}$ are respectively defined in Eqs~\eqref{eq:A} and \eqref{eq:src}.
We explain this relationship utilizing an example of $m=2$.
When $m=2$, the Hermitian $A$ and the quantum state $\Ket{b}$ can be written in the matrix form, as follows:
\begin{equation}
	A = \begin{bmatrix}
		\varepsilon K_{1} + \varepsilon K_{2} & O & O & O \\
		O & \varepsilon K_{1} + K_{2} & O & O \\
		O & O & K_{1} + \varepsilon K_{2} & O \\
		O & O & O & K_{1} + K_{2} \\
	\end{bmatrix},
\end{equation}
\begin{equation}
	\Ket{b} = \dfrac{1}{2} \begin{bmatrix}
		\bs{F} \\
		\bs{F} \\
		\bs{F} \\
		\bs{F} \\
	\end{bmatrix}.
\end{equation}
Therefore, solving Eq.~\eqref{eq:gov_q} with respect to $\Ket{\Psi}$ yields
\begin{align}
	\Ket{\Psi} & = A^{-1} \Ket{b}
	= \dfrac{1}{2} \begin{bmatrix}
		(\varepsilon K_{1} + \varepsilon K_{2})^{-1} \bs{F} \\
		(\varepsilon K_{1} + K_{2})^{-1} \bs{F} \\
		(K_{1} + \varepsilon K_{2})^{-1} \bs{F} \\
		(K_{1} + K_{2})^{-1} \bs{F} \\
	\end{bmatrix}
	 = \dfrac{1}{2} \begin{bmatrix}
		\bs{U}(0,0) \\
		\bs{U}(0,1) \\
		\bs{U}(1,0) \\
		\bs{U}(1,1) \\
	\end{bmatrix},
\end{align}
where $U(x_1, x_2)$ represents the solution of Eq.~\eqref{eq:gov}.
That is, the solution of Eq.~\eqref{eq:gov_q} includes the temperature vectors in all possible configurations.

To solve Eq.~\eqref{eq:gov_q}, we now introduce an objective function to be minimized, as follows:
\begin{align}
      E(\Ket{\Psi}):= \dfrac{1}{2} \Braket{\Psi | A | \Psi} - \sqrt{\Braket{\Psi \Ket{b} \Bra{b} \Psi}}. \label{eq:energy}
\end{align}
Since the $A$ is positive definite, owing to the positive definiteness of the stiffness matrix $K_j$, this objective function has a unique minimum, and the stationary condition, i.e., $dE/d\ket{\Psi}$ requiring that the gradient at the minimum equals to $0$ yields
\begin{align}
      A \Ket{\Psi} = \dfrac{\Braket{b | \Psi}}{\| \Braket{b | \Psi} \|} \Ket{b},
\end{align}
which implies that Eq.~\eqref{eq:gov_q} holds up to a global phase of $\Ket{\Psi}$ at the minimum of Eq.~\eqref{eq:energy}.
Because the global phase in quantum states has no relevance, Eq.~\eqref{eq:gov_q} may be solved by minimizing $E$ in Eq.~\eqref{eq:energy} with respect to $\Ket{\Psi}$.
It should noted here that the norm of the vector $\Ket{\Psi}$ is not necessarily $1$ because the operator $A$ is not unitary.
This means that $\Ket{\Psi}$ is not a quantum state.
Thus, to deal with the vector $\Ket{\Psi}$ in quantum computers, we introduce a quantum state $\Ket{\psi (\bs{\theta})}$ and a parameter $r \in \mathbb{R}$ and define $\Ket{\Psi}:= r \Ket{\psi(\bs{\theta})}$.
Because the quantum state $\Ket{\psi (\bs{\theta})}$ is parametrized by parameters $\bs{\theta} \in \mathbb{R}^p$ with $p$ being the number of parameters, we can replace the problem of minimizing $E$ in Eq.~\eqref{eq:energy} with respect to $\Ket{\Psi}$ by a problem of finding parameters $(\bs{\theta}, r) \in \mathbb{R}^p \times \mathbb{R}$ that minimize $E'$ defined as
\begin{align}
      E' (\bs{\theta}, r):= \dfrac{1}{2} r^2\Braket{\psi (\bs{\theta}) | A | \psi (\bs{\theta})} - r\sqrt{\Braket{\psi (\bs{\theta}) \Ket{b} \Bra{b} \psi (\bs{\theta})}}. \label{eq:energy2}
\end{align}
The minimizer of $r$ can calculated analytically due to the parabolic nature of $E'$ with respect to $r$, be obtained as
\begin{align}
      r^\ast (\bs{\theta}) = \dfrac{\sqrt{\Braket{\psi(\bs{\theta})\Ket{b} \Bra{b} \psi(\bs{\theta})}}}{\Braket{\psi (\bs{\theta}) | A | \psi (\bs{\theta})}}. \label{eq:r}
\end{align}
Then, the objective function to be minimized with respect to $\bs{\theta}$ is derived, by substituting the minimizer of $r$ in Eq.~\eqref{eq:r} into Eq.~\eqref{eq:energy2}, as follows:
\begin{align}
	& E' (\bs{\theta}, r^\ast(\bs{\theta})) = -\dfrac{1}{2} \dfrac{\Braket{\psi(\bs{\theta})\Ket{b} \Bra{b} \psi(\bs{\theta})}}{\Braket{\psi (\bs{\theta}) | A | \psi (\bs{\theta})}}.
\end{align}
Omitting the constant $1/2$ results in an objective function to be minimized for solving the governing equation~\eqref{eq:gov} defined as
\begin{align}
	& F_u (\bs{\theta}) := -\dfrac{\Braket{\psi(\bs{\theta})\Ket{b} \Bra{b} \psi(\bs{\theta})}}{\Braket{\psi (\bs{\theta}) | A | \psi (\bs{\theta})}}. \label{eq:obj_u}
\end{align}
Consequently, the optimization problem in Eq.\eqref{eq:opt_u} in the Result\nameref{sec:results} section is derived as
\begin{align}
	& \min_{\bs{\theta}} \quad F_u (\bs{\theta}) = -\dfrac{\Braket{\psi(\bs{\theta})\Ket{b} \Bra{b} \psi(\bs{\theta})}}{\Braket{\psi (\bs{\theta}) | A | \psi (\bs{\theta})}}. \label{eq:opt_u_2}
\end{align}

\subsubsection*{Solving structural optimization}
Next, we formulate an another optimization problem for solving the original optimization problem in Eqs.~\eqref{eq:obj} and \eqref{eq:gov} by quantum computing.
In particular, we consider the objective function expressed written in the form
\begin{align}
      \mathcal{L}(U(\bs{x})) = \bs{U}^\dagger O \bs{U}, \label{eq:obj_specific_2}
\end{align}
where $O$ is a Hermitian matrix and $\dagger$ represents the Hermitian transpose.
The purpose of structural optimization is to find the structure $\bs{x}$ that minimize the objective function $\mathcal{L}$.

Let $\bs{\theta}^\ast$ denote the optimal parameters of the problem in Eq.~\eqref{eq:opt_u_2}.
Then, the weighted sum of the objective function values for all conceivable structures can be written as
\begin{align}
    \sum_{\bs{x} \in \{0,1\}} P(\bs{x}) \bs{U}(\bs{x})^\dagger O \bs{U}(\bs{x}) &= r^2 \sum_{\bs{x} \in \{0,1\}} P(\bs{x}) \Braket{\psi (\bs{\theta}^\ast) | \left( \Ket{\bs{x}} \Bra{\bs{x}} \otimes O \right) | \psi (\bs{\theta}^\ast) } \nm \\
    &= r^2 \Braket{\psi (\bs{\theta}^\ast) | \left( \rho \otimes O \right) | \psi (\bs{\theta}^\ast) }, \label{eq:obj_q_mix}
\end{align}
where the weighting coefficient for each structure corresponds to its probability $P(\bs{x})$ and 
\begin{align}
    \rho := \sum_{\bs{x} \in \{0,1\}} P(\bs{x}) \Ket{\bs{x}} \Bra{\bs{x}}. \label{eq:rho}
\end{align}
When the likelihood of the structure with the least objective function value is highest, the weighed sum of the objective function achieves the minimum.
Consequently, minimizing Eq.~\eqref{eq:obj_q_mix} with respect to the probability $P (\bs{x})$ yields the probability distribution of $P (\bs{x}^\ast) = 1$ where $\bs{x}^\ast$ is the optimal structure.
Thus, we now parametrize the probability $P(\bs{x})$, by introducing a parametrized quantum state $\ket{\phi (\bs{\eta})}$ on $m$ qubits, as follows:
\begin{align}
    P_\eta (\bs{x}) := \left| \Braket{\bs{x} | \phi (\bs{\eta})} \right|^2, \label{eq:probability}
\end{align}
which subsequently yields a parametrized mixed state $\rho (\bs{\phi})$ as
\begin{align}
    \rho (\bs{\eta}) := \sum_{\bs{x} \in \{0,1\}^m} P_\eta (\bs{x}) \Ket{\bs{x}} \Bra{\bs{x}}.
\end{align}
Omitting $r^2$, which is fixed after the first optimization step in Eq.~\eqref{eq:opt_u} is completed and is constant with regard to $\bs{\eta}$, we may eventually state the structural optimization problem, as follows:
\begin{align}
    & \min_{\bs{\eta}} \quad F_s(\bs{\eta}, \bs{\theta}^\ast), \label{eq:opt_s_2}
\end{align}
where
\begin{align}
    F_s(\bs{\eta}, \bs{\theta}) := \Braket{\psi (\bs{\theta}) | \left( \rho (\bs{\eta}) \otimes O \right) | \psi (\bs{\theta}) }. \label{eq:obj_s}
\end{align}
When the optimized parameters $\bs{\eta}^\ast$ is obtained, measuring the quantum state $\Ket{\phi (\eta^\ast)}$ produces the optimal structure $\bs{x}^\ast$ with a probability of $1$ if the entire operation is completed successfully.
It is remarkable that the proposed method requires no amplitude estimation with respect to $\Ket{\psi (\theta)}$.
That is, the temperature obtained in the first stage is employed in the second step for structural optimization while it is embedded in the amplitude of a quantum state.
The bit string representing the optimal structure is the end result of the method, and this information is easily acquired from the quantum state.

\subsection*{Implementation}
\subsubsection*{Parametrized quantum circuit}
\begin{figure}[t]
\centering
\subfloat[][]{\includegraphics[width=0.4\linewidth]{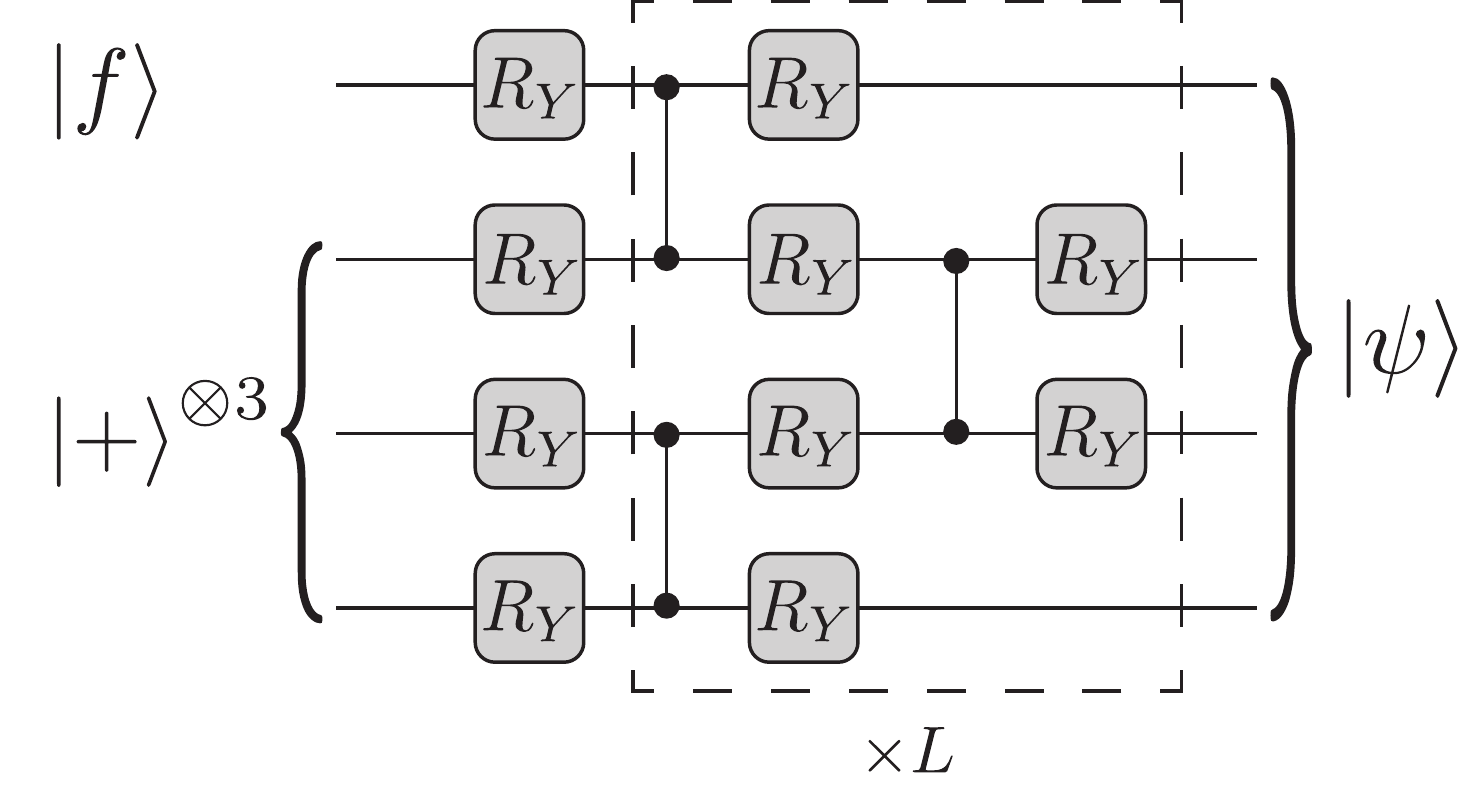} \label{subfig:ansatz_theta}} \quad
\subfloat[][]{\includegraphics[width=0.4\linewidth]{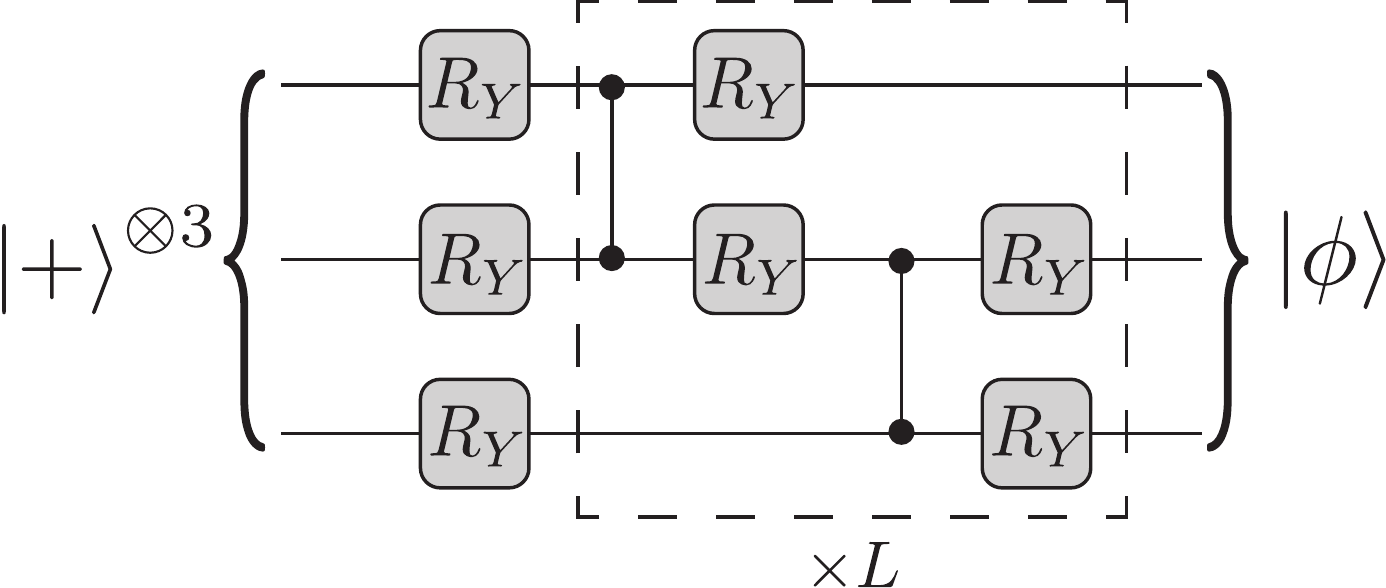} \label{subfig:ansatz_phi}}
\caption[]{Alternating layered ansatz used for the three-edge design problem. \subref{subfig:ansatz_theta} Ansatz for generating $\Ket{\psi(\theta)}$. \subref{subfig:ansatz_phi} Ansatz for generating $\Ket{\phi(\eta)}$. \label{fig:ansatz}}
\end{figure}
To produce two kinds of quantum states, $\Ket{\psi (\bs{\theta})}$ and $\Ket{\phi (\bs{\eta})}$,
we used the alternating layered ansatz~\cite{Cerezo2021cost}, illustrated in Fig.~\ref{fig:ansatz}.
Specifically, the number of layers $L$ was set to $2$ for generating $\Ket{\psi (\theta)}$, while it was set to $1$ for $\Ket{\phi (\eta)}$.
Although it is well-known that this form of parametrized quantum circuits suffers from the exponentially vanishing gradients unless observables are local~\cite{Cerezo2021cost}, we can mitigate this problem, to some extent, by carefully choosing initial parameters.
Let $U_\theta$ denote a parametrized quantum circuit.
We now parametrize the quantum state $\ket{\psi (\bs{\theta})}$ using the parametrized quantum circuit $U_\theta$ and the quantum state $\ket{b}$, as follows:
\begin{align}
    \Ket{\psi (\bs{\theta})} = U_\theta \Ket{b}.
\end{align}
Setting all parameters $\bs{\theta}$ to $0$ and the number of layers $L$ to an even number avoids barren plateaus at least for the initial parameters, but not for the entire optimization procedure.
Detailed discussion about this initialization is provided in the supplementary material.

For the initialization of parameters $\bs{\eta}$, which are for expressing the quantum superposition of structures, we employ a similar strategy.
Denoting a parametrized quantum circuit by $U_\eta$, we parametrize the quantum state $\ket{\phi (\bs{\eta})}$, as follows:
\begin{align}
    \Ket{\phi (\bs{\eta})} = U_\eta \Ket{+}^{\otimes m}.
\end{align}
We set all initial parameters $\bs{\eta}$ to $0$, which means that all possible structures are in quantum superposition with a uniform probability.
Note that the number of layers $L$ for the parametrized quantum circuit $U_\eta$ is not necessary an even number, different from the case of $U_\theta$.
When it is an odd number, the entangler, CZ gates in the ansatz, causes the phase difference among the structures in quantum superposition.

\subsubsection*{Algorithm}
Based on the formulation in the previous sections, we now discuss the algorithm for quantum topology optimization.

\begin{enumerate}
    \item Initialize both parameters $\bs{\theta}$ and $\bs{\eta}$ to $\bs{0}$. The parameter initialization strategy is described in the supplementary information.
    \item Solve the optimization problem in Eq.~\eqref{eq:opt_u} for obtaining state fields. Let $\bar{\bs{\theta}}$ denote the obtained solution. The parameters which give the lowest objective function value in the optimization history are output as the solution $\bar{\bs{\theta}}$.
    \item Solve the optimization problem in Eq.~\eqref{eq:opt_s} for amplifying the probability of the optimum structures, using the solution $\bar{\bs{\theta}}$ in Step 2 as $\bs{\theta}^\ast$. Let $\bar{\bs{\eta}}$ denote the obtained solutions. Again, the parameters which give the lowest objective function value in the optimization history are output as the solution $\bar{\bs{\eta}}$.
    \item Prepare the quantum state $\ket{\varphi (\bar{\bs{\eta}})}$ and measure it in the computational basis. The structure corresponding to the bit string observed with the highest probability is output as the optimal structure.
\end{enumerate}

The objective functions are assessed using quantum computing expectation evaluations during the optimization method.
The expectation evaluations are performed based on the extended Bell measurement~\cite{Kondo2022computationally} and the inversion test~\cite{Ruan2021quantum}.
The supplemental information provided a detailed explanation.

\bibliography{ref}

\section*{Acknowledgments}
This work was partially supported by UTokyo Quantum Initiative.
We thank Prof. Imoto for insightful comments.

\section*{Author contributions statement}
Y. S. and R. K. conceived the formulation, Y. S. developed the codes and performed the experiments, S. Koide and S. Kajita analysed the results.
All authors contributed to the discussion and reviewed the manuscript.

\section*{Competing interests}
The authors declare no conflicts of interest associated with this manuscript.

\end{document}

\maketitle

\section*{Extension of the system size to have the dimension of $2^n$} \label{sec:gov_q_ex}
To handle the governing equation in a $n$-qubits quantum system, we extended the governing equation in Eq.~\eqref{eq:gov} to have the dimension of $2^n$ where $n:= \lceil \log_2 N \rceil$.
We introduce the extended matrix $\overline{K}_j$ of the element stiffness matrix $K_j$ defined as
\begin{align}
    \overline{K}_j := \begin{bmatrix}
    K_j & O \\
    O & I_{2^n-N}
    \end{bmatrix},
\end{align}
where $I_{2^n-N}$ is the identity matrix with the size of $(2^n-N) \times (2^n-N)$.
Similarly, we also introduce the extended vector $\overline{\bs{F}}$ of the input vector $\bs{F}$ to have the size of $(2^n-N)$, which is defined as
\begin{align}
    \overline{\bs{F}} := \begin{bmatrix}
    \bs{F} \\
    \bs{0}_{2^n-N}
    \end{bmatrix},
\end{align}
where $\bs{0}_{2^n-N}$ is the zero vector with the size of $2^n-N$.
Then, the governing equation in Eq.~\eqref{eq:gov} is extended as
\begin{align}
    \left( \sum_{j=1}^m \overline{K}_j \left( (1-\varepsilon) x_j + \varepsilon \right) \right) \overline{\bs{U}} = \overline{\bs{F}}. \label{eq:exgov}
\end{align}
Clearly, the extended solution vector $\overline{\bs{U}}$ consists of 
\begin{align}
\begin{bmatrix}
\bs{U} \\ 
\bs{0}_{2^n-N}
\end{bmatrix}.
\end{align}
Therefore, we can solve this extended governing equation with the size of $2^n$ instead of the original one with the size of $N$.

\section*{Expectation evaluation for calculating objective function} \label{sec:exp}
Here, we explain how to evaluate the expectation values in the objective function, i.e., the expectations of $A$ and $\Ket{b}\Bra{b}$ in Eq.~\eqref{eq:opt_u}, and $\rho (\eta) \otimes O$ in Eq.~\eqref{eq:opt_s}.

\subsection*{Expectation of $A$} \label{sec:exp_A}
The operator $A$ is written as 
\begin{equation}
	A := \sum_{j=1}^m \dfrac{1}{2}\left( (1 + \varepsilon) I^{\otimes m} - (1 - \varepsilon) Z_j \right) \otimes K_j. \label{s:eq:A}
\end{equation}
Based on the Fourier's law, the element stiffness matrix $K_j$ of the $j$-th edge, connecting the $i$-th node $v_i$ and the $i'$-th node $v_{i'}$, which has the thermal conductivity $\lambda$ and the length $l$, is given as
\begin{align}
    K_j = \dfrac{\lambda}{l} \left( \Ket{v_i}\Bra{v_i} + \Ket{v_{i'}}\Bra{v_{i'}} - \Ket{v_i}\Bra{v_{i'}} - \Ket{v_{i'}}\Bra{v_i} \right).
\end{align}
Based on the concept of the extended Bell measurement (XBM)~\cite{kondo2022}, we reformulate $K_j$ as
\begin{align}
    K_j &= \dfrac{\lambda}{l} \left( \dfrac{\Ket{v_i} + \Ket{v_{i'}}}{\sqrt{2}}\dfrac{\Bra{v_i} + \Bra{v_{i'}}}{\sqrt{2}} + \dfrac{\Ket{v_i} - \Ket{v_{i'}}}{\sqrt{2}}\dfrac{\Bra{v_i} - \Bra{v_{i'}}}{\sqrt{2}} \right) \\
    &= \dfrac{\lambda}{l} \left( M_{+} \Ket{v_i} \Bra{v_i} M_{+}^\dagger + M_{-} \Ket{v_i} \Bra{v_i} M_{-}^\dagger \right),
\end{align}
where $M_{\pm}$ is the unitary satisfying
\begin{align}
    M_{+} \Ket{v_i} &= \dfrac{\Ket{v_i} + \Ket{v_{i'}}}{\sqrt{2}} \nm \\
    M_{-} \Ket{v_i} &= \dfrac{\Ket{v_i} - \Ket{v_{i'}}}{\sqrt{2}},
\end{align}
that is, the unitary generating the Bell state of $\Ket{v_i}$ and $\Ket{v_{i'}}$.
Then, the operator $A$ is rewritten, as follows:
\begin{align}
	A =& \sum_{j=1}^m \dfrac{\lambda}{2l} \left( (1 + \varepsilon) I^{\otimes m} - (1 - \varepsilon) Z_j \right) \otimes  \left( M_{+} \Ket{v_i} \Bra{v_i} M_{+}^\dagger + M_{-} \Ket{v_i} \Bra{v_i} M_{-}^\dagger \right) \nm \\
	=& \dfrac{\lambda}{2l} (1 + \varepsilon) \sum_{j=1}^m I^{\otimes m} \otimes M_{+} \Ket{v_i} \Bra{v_i} M_{+}^\dagger + \dfrac{\lambda}{2l} (1 + \varepsilon) \sum_{j=1}^m I^{\otimes m} \otimes M_{-} \Ket{v_i} \Bra{v_i} M_{-}^\dagger \nm \\
	&- \dfrac{\lambda}{2l} (1 - \varepsilon) \sum_{j=1}^m Z_j  \otimes M_{+} \Ket{v_i} \Bra{v_i} M_{+}^\dagger - \dfrac{\lambda}{2l} (1 - \varepsilon) \sum_{j=1}^m Z_j  \otimes M_{-} \Ket{v_i} \Bra{v_i} M_{-}^\dagger.
\end{align}
The expectations of these four terms can respectively be evaluated using the Pauli measurement for the leftmost $m$ qubits and the XBM for the rightmost $n$ qubits.

\subsection*{Expectation of $\Ket{b}\Bra{b}$}
Let $U_f$ denote the unitary satisfying
\begin{equation}
    \Ket{f} = U_f \Ket{0}^{\otimes n}.
\end{equation}
The quantum state $\Ket{b} \Bra{b}$ can then be represented, based on Eq.~\eqref{eq:src}, as follows:
\begin{align}
    \Ket{b} \Bra{b} &= \Ket{+}^{\otimes m}\Bra{+}^{\otimes m} \otimes \Ket{f} \Bra{f} = \left( H^{\otimes m} \otimes U_f \right) \Ket{0}^{\otimes (m+n)} \Bra{0}^{\otimes (m+n)} \left( H^{\otimes m} \otimes U_f^\dagger \right).
\end{align}
Therefore, the expectation of $\Ket{b} \Bra{b}$ can be evaluated by the inversion test.

\subsection*{Expectation of $\rho (\eta) \otimes O$}
Based on the definitions of the state $\rho(\eta)$ in Eq.~\eqref{eq:rho} and the operator $O$ in Eq.~\eqref{eq:opt_s_o}, the operator $\rho (\eta) \otimes O$ can be expressed as
\begin{align}
    \rho (\eta) \otimes O &= \sum_{\bs{x} \in \{0,1\}} P_\eta (\bs{x}) \Ket{\bs{x}} \Bra{\bs{x}} \otimes \Ket{v_\text{protected}} \Bra{v_\text{protected}} \nm \\
    &= \sum_{\bs{x} \in \{0,1\}} P_\eta (\bs{x}) \Ket{\bs{x}} \Bra{\bs{x}} \otimes \left( \prod_{i=1}^n U_i \right) \Ket{0}^{\otimes n} \Bra{0}^{\otimes n} \left( \prod_{i=1}^n U_i^\dagger \right),
\end{align}
where
\begin{equation}
    U_j = \begin{cases}
    X & \text{if the }j\text{-th digit of the bit representation of }v_\text{protected}\text{ is 1} \\
    I & \text{otherwise}.
    \end{cases}
\end{equation}
Because $\Ket{\bs{x}}\Bra{\bs{x}}$ for $\bs{x} \in \{0, 1 \}^m$ are the projection operators, the expectation of $\rho (\eta) \otimes O$ can be evaluated by the projective measurement for the lestmost $m$ qubits and the inversion test~\cite{ruan2021} for the rightmost $n$ qubits.

\section*{Parameter Initialization}
The parameter initialization strategy for both problems in Eqs.~\eqref{eq:opt_u} and \eqref{eq:opt_s} is discussed.
First, we consider the initialization of parameters $\bs{\theta}$, which are introduced for expressing the state field.
In the present study, since we use the alternating layered ansatz and both the observables $A$ and $\ket{b} \bra{b}$ are global, the exponentially vanishing gradient problems, also called barren plateaus, will arise for the randomly initialized parameters of a parametrized quantum circuit~\cite{cerezo2021}.
To alleviate this, we use a deterministic initialization of parameters, which is the similar as the previous research~\cite{grant2019}.
Since the matrix $A$ in Eq.~\eqref{eq:A} is positive definite, its expectation value is always positive.
On the other hand, because the expectation of the operator $\ket{b} \bra{b}$ corresponds to the fidelity between $\ket{\psi (\bs{\theta})}$ and $\Ket{b}$, it will exponentially be small as the number of qubits increases when the $\ket{\psi (\bs{\theta})}$ is prepared by random parameters.
This implies that the cost function value becomes $0$ almost everywhere in the parameter space, which leads to barren plateaus.
Actually, the gradient of $F_u$ with respect to $i$-th parameter $\theta_i$ is derived as
\begin{align}
    \dfrac{\partial F_u}{\partial \theta_i} = - \dfrac{ \text{Re} \left( \Braket{\psi_{,i}(\bs{\theta})\Ket{b} \Bra{b} \psi(\bs{\theta})} \right)}{\Braket{\psi (\bs{\theta}) | A | \psi (\bs{\theta})}} + \dfrac{\Braket{\psi(\bs{\theta})\Ket{b} \Bra{b} \psi(\bs{\theta})} \text{Re} \left( \Braket{\psi_{,i} (\bs{\theta}) | A | \psi (\bs{\theta})} \right) }{\Braket{\psi (\bs{\theta}) | A | \psi (\bs{\theta})}^2},
\end{align}
where $\ket{\psi_{,i}(\bs{\theta})}$ represents $\partial \ket{\psi(\bs{\theta})} / \partial \theta_i$.
This equation indicates that the gradients become $0$ when the fidelity $| \braket{b | \psi (\bs{\theta})} |^2$ is $0$.
Therefore, we consider the parameter initialization which ensures that the fidelity $| \braket{b | \psi (\bs{\theta})} |^2$ is not $0$.
Let $U_\theta$ denote a parametrized quantum circuit.
We now parametrize the quantum state $\ket{\psi (\bs{\theta})}$ using the parametrized quantum circuit $U_\theta$ and the quantum state $\ket{b}$, as follows:
\begin{align}
    \Ket{\psi (\bs{\theta})} = U_\theta \Ket{b}.
\end{align}
Setting parameters $\bs{\theta}$ so that $U_\theta$ could be the identity makes the parametrized quantum state $\ket{\psi (\bs{\theta})}$ coincide to the quantum state $\ket{b}$, which means that the fidelity $| \braket{b | \psi (\bs{\theta})} |^2$ becomes $1$.
In using the alternating layered ansatz, this can be realized by setting all parameters $\bs{\theta}$ to $0$ and the number of layers $L$ to an even number.
Consequently, such initialization avoids barren plateaus at least for the initial parameters although it does not for the entire optimization procedure.

As for the initialization of parameters $\bs{\eta}$, which are for expressing the quantum superposition of structures, we introduce the similar strategy.
Denoting a parametrized quantum circuit by $U_\eta$, we parametrize the quantum state $\ket{\phi (\bs{\eta})}$, as follows:
\begin{align}
    \Ket{\phi (\bs{\eta})} = U_\eta \Ket{+}^{\otimes m}.
\end{align}
In the present study, we set all initial parameters $\bs{\eta}$ to $0$, which means that all possible structures are in quantum superposition with a uniform probability.
Note that the number of layers $L$ for the parametrized quantum circuit $U_\eta$ is not necessary an even number, different from the case of $U_\theta$.
When it is an odd number, the phase difference arises among the structures in quantum superposition due to the entangler, CZ gates in the ansatz.

\section*{Supplementary of numerical experiments}
\subsection*{Five-edge-design-problem}
\begin{figure}[t]
\centering
\subfloat[][]{\includegraphics[width=0.6\linewidth]{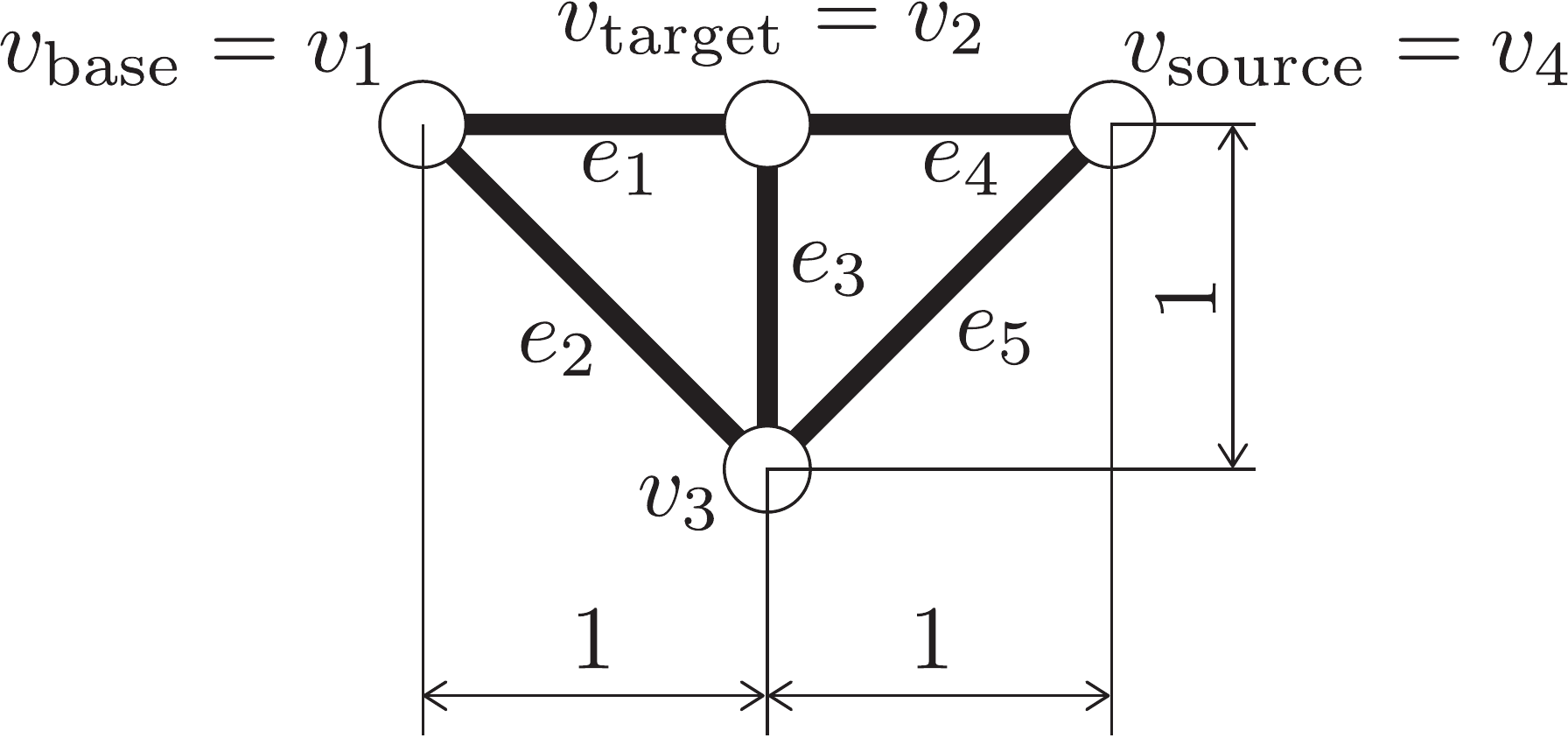} \label{subfig:5edges_setting}}
\subfloat[][]{\includegraphics[width=0.3\linewidth]{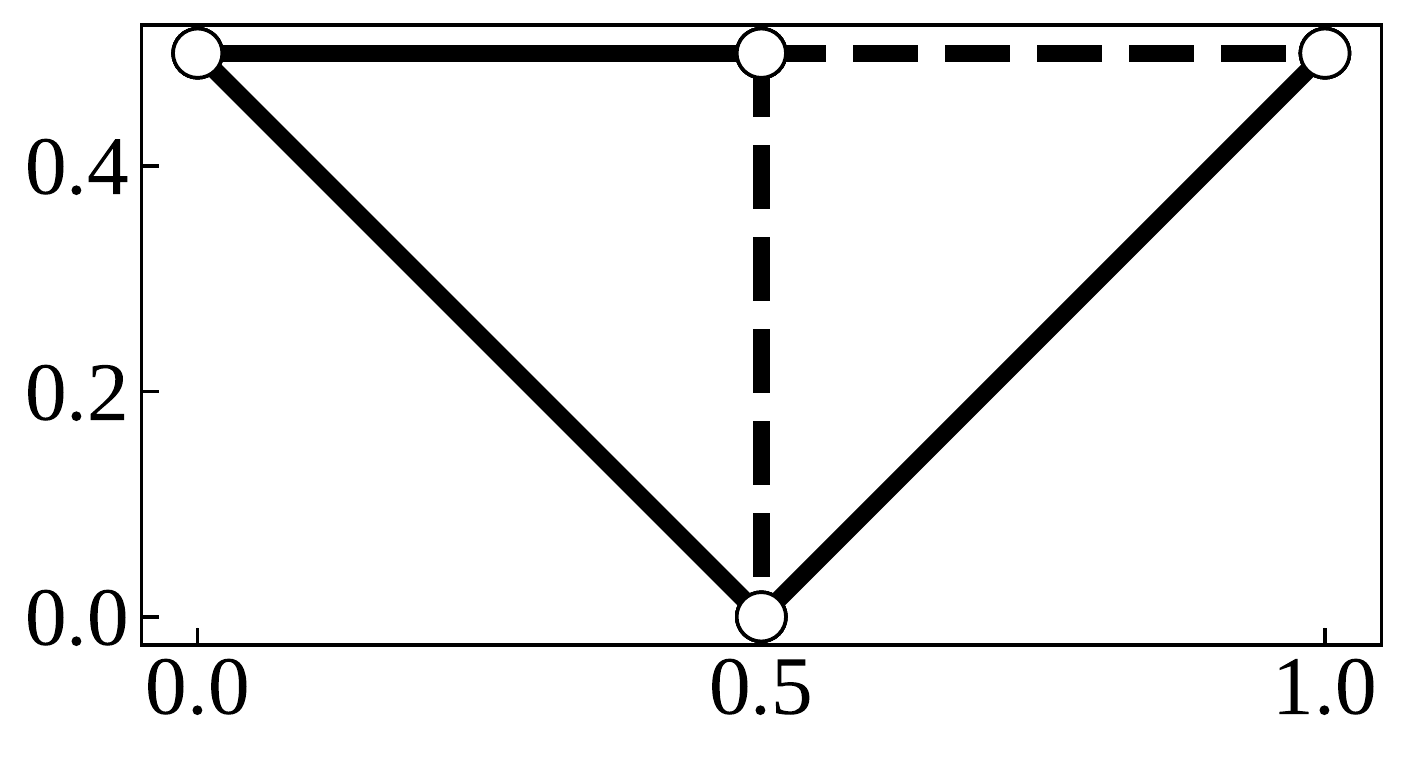} \label{subfig:5edges_optstr}}
\caption{The five-edge-design problem. (a) Setting of the problem. (b) The structure corresponding to the bit string 11001. The bold edges represents the material with bigger thermal conductivity, i.e., $x_j=1$, while the dashed edge represents the material with smaller thermal conductivity, i.e., $x_j=0$. This structure coincides to the exact optimal structure obtained by the brute-force search.}
\label{fig:5edges_prob}
\end{figure}
For supplementary results, we provide the results of the design problem of five edges, whose settings were illusrated in Fig.~\ref{fig:5edges_prob} \subref{subfig:5edges_setting}.
The objective function to be minimized is the temperature on the node $v_\text{protected}$ under the condition that the temperature was fixed to $0$ on the node $v_\text{base}$, and the heat source was applied on the node $v_\text{source}$.
The optimal structure which is obtained by classical brute-force search is shown in Fig.~\ref{fig:5edges_prob} \ref{subfig:5edges_optstr}.
The number of layers for the alternating layered ansatz was set to $6$ for generating the state $\Ket{\psi (\theta)}$ and $2$ for the state $\Ket{\phi (\eta)}$.

\subsubsection*{Statevector simulation}
%%%%%%%%%%%%%%%%%%%%%%%%%%%%%%%%%%%%%%%%%%%%%%%%%%%%
\begin{figure}[t]
	\centering
	\subfloat[][]{\includegraphics[width=0.7\linewidth]{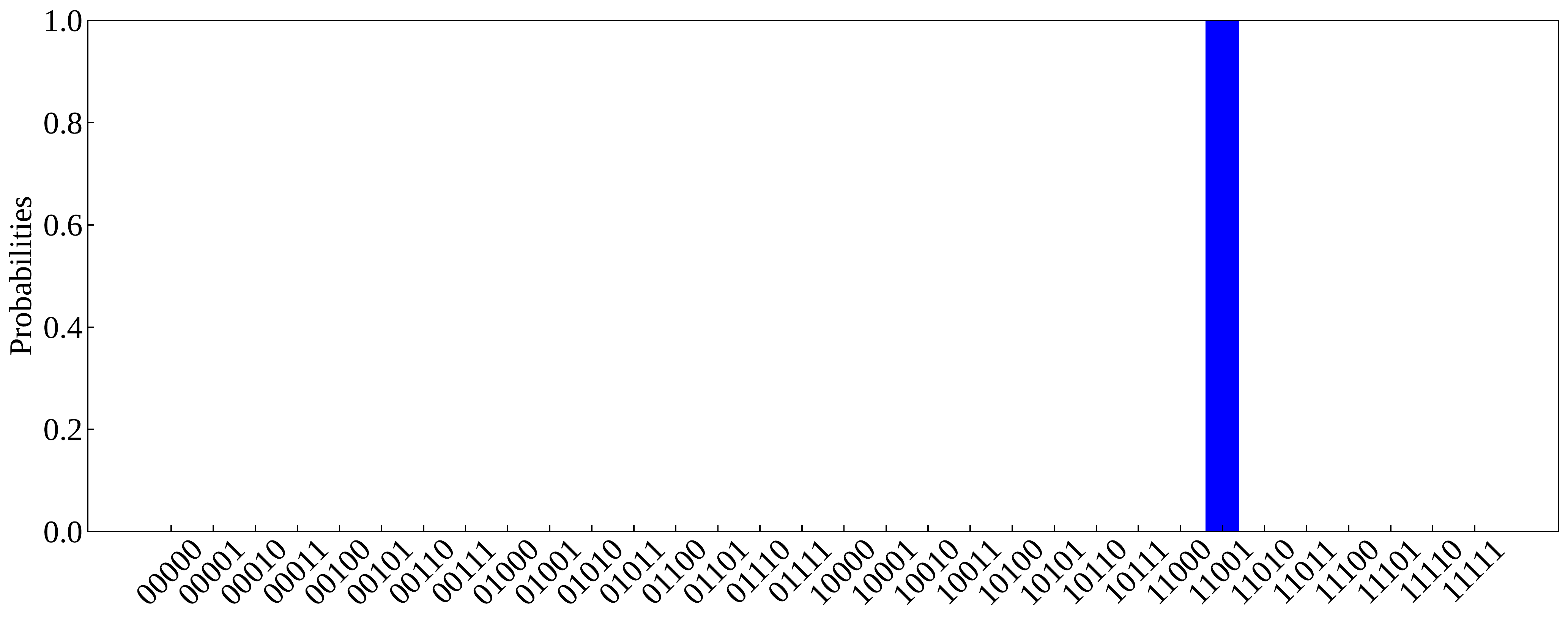} \label{subfig:5edges_probability}} \\
	\subfloat[][]{\includegraphics[width=0.7\linewidth]{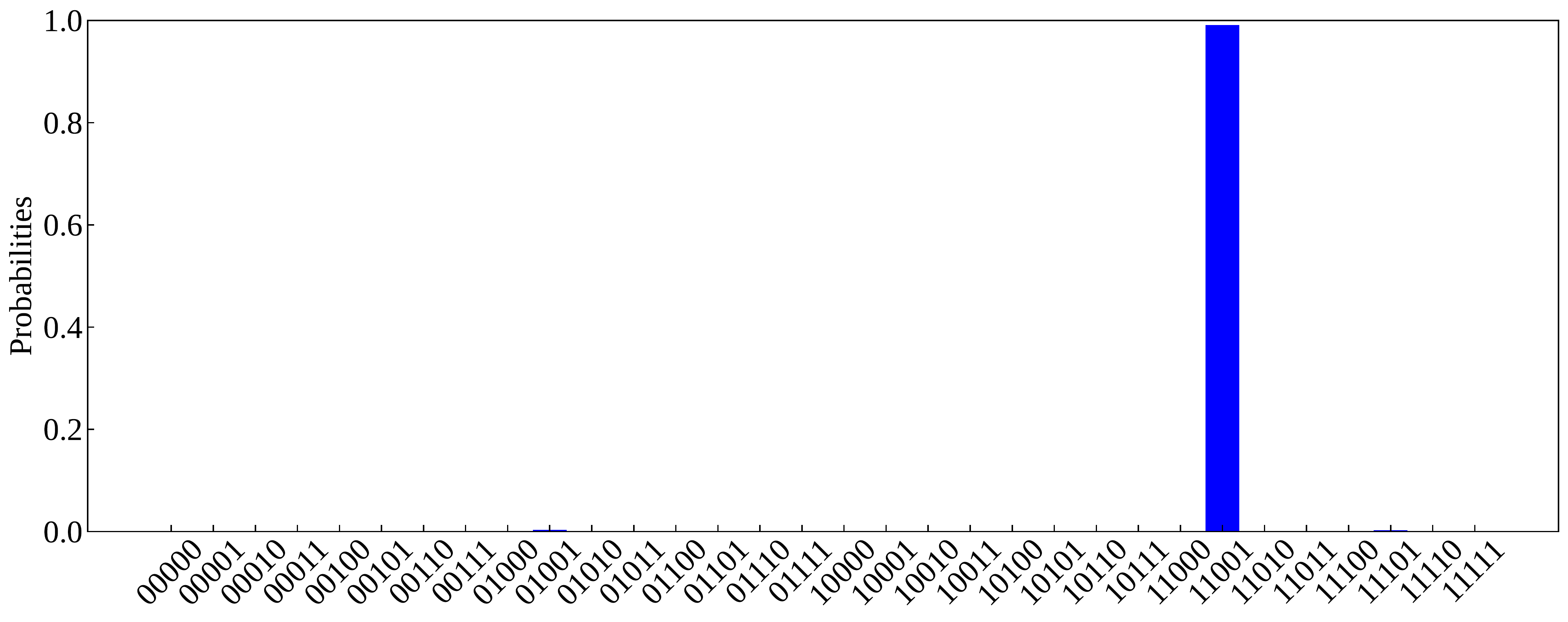} \label{subfig:5edges_shots_probability}} \\
	\subfloat[][]{\includegraphics[width=0.35\linewidth]{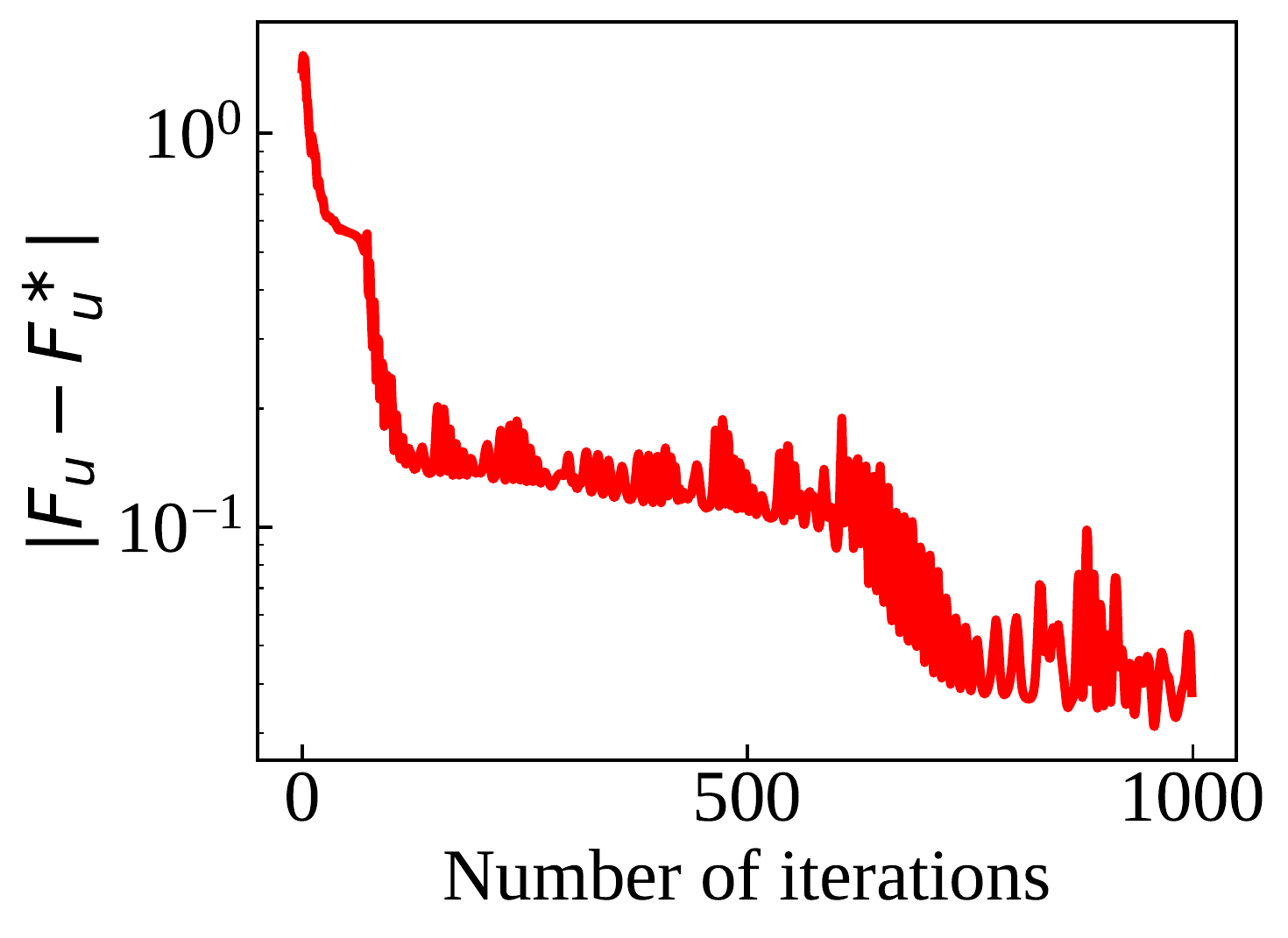} \label{subfig:5edges_history_state}} \quad
	\subfloat[][]{\includegraphics[width=0.35\linewidth]{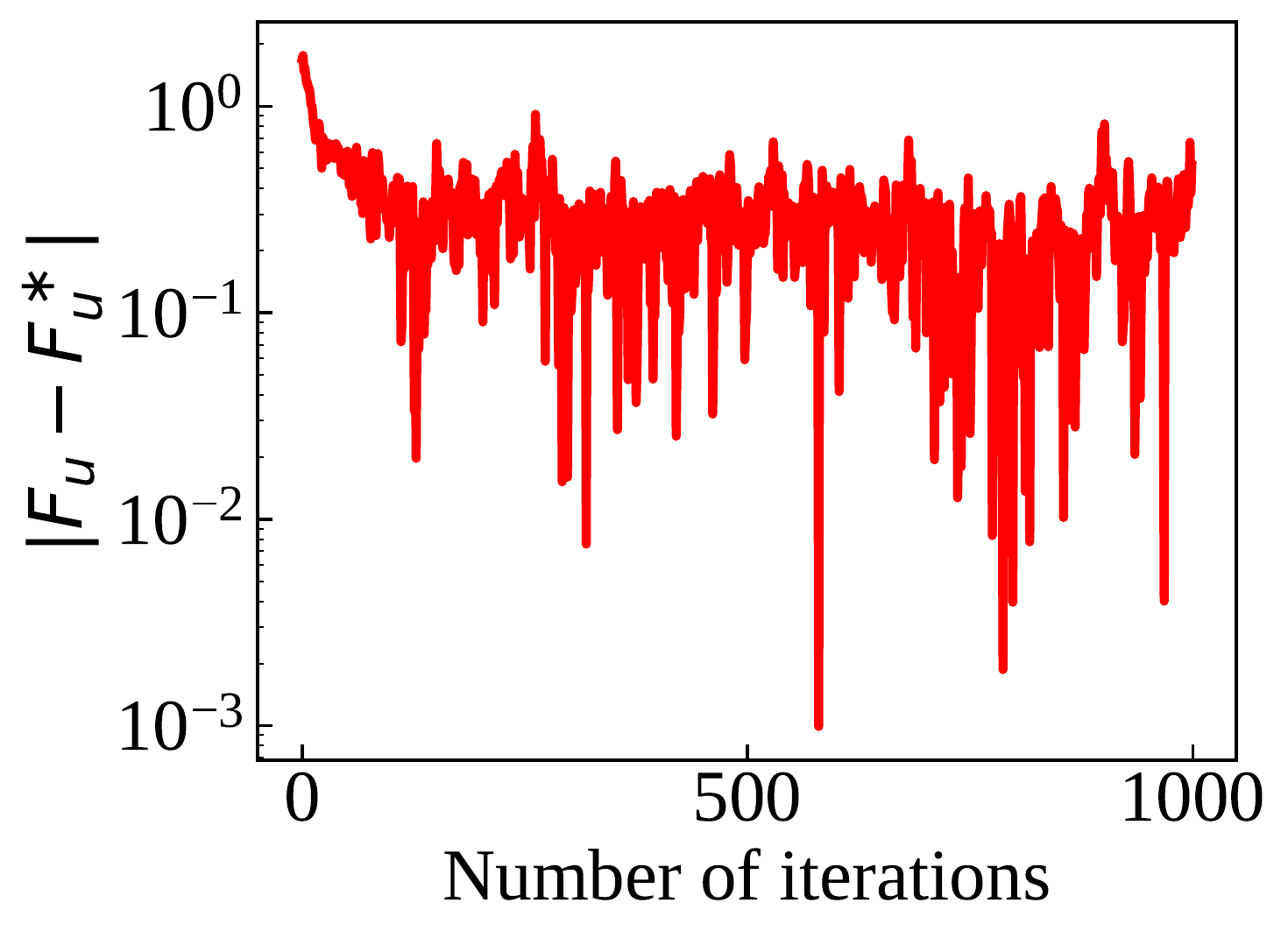} \label{subfig:5edges_shots_history_state}} \\
	\subfloat[][]{\includegraphics[width=0.35\linewidth]{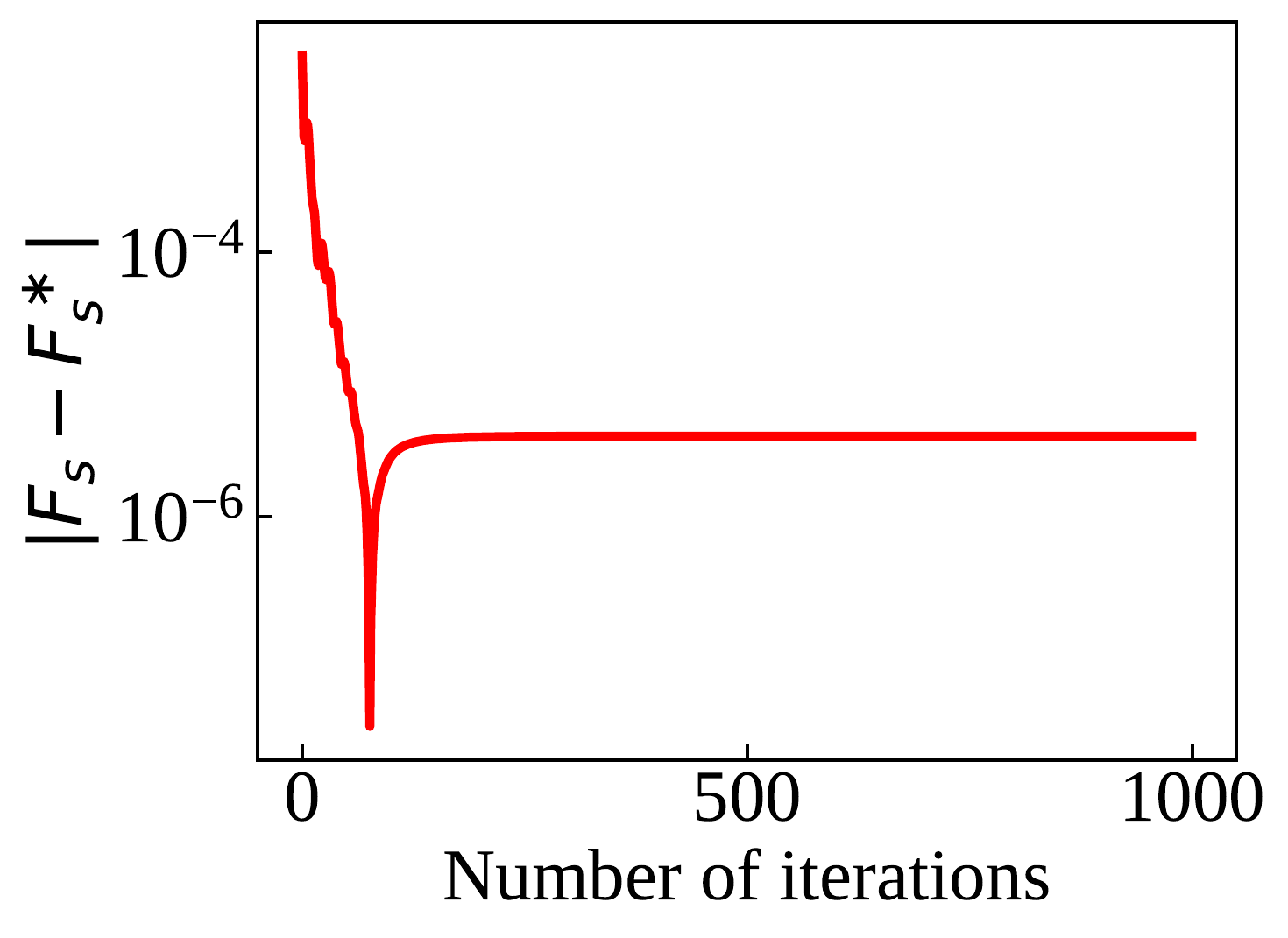} \label{subfig:5edges_history_struct}} \quad
	\subfloat[][]{\includegraphics[width=0.35\linewidth]{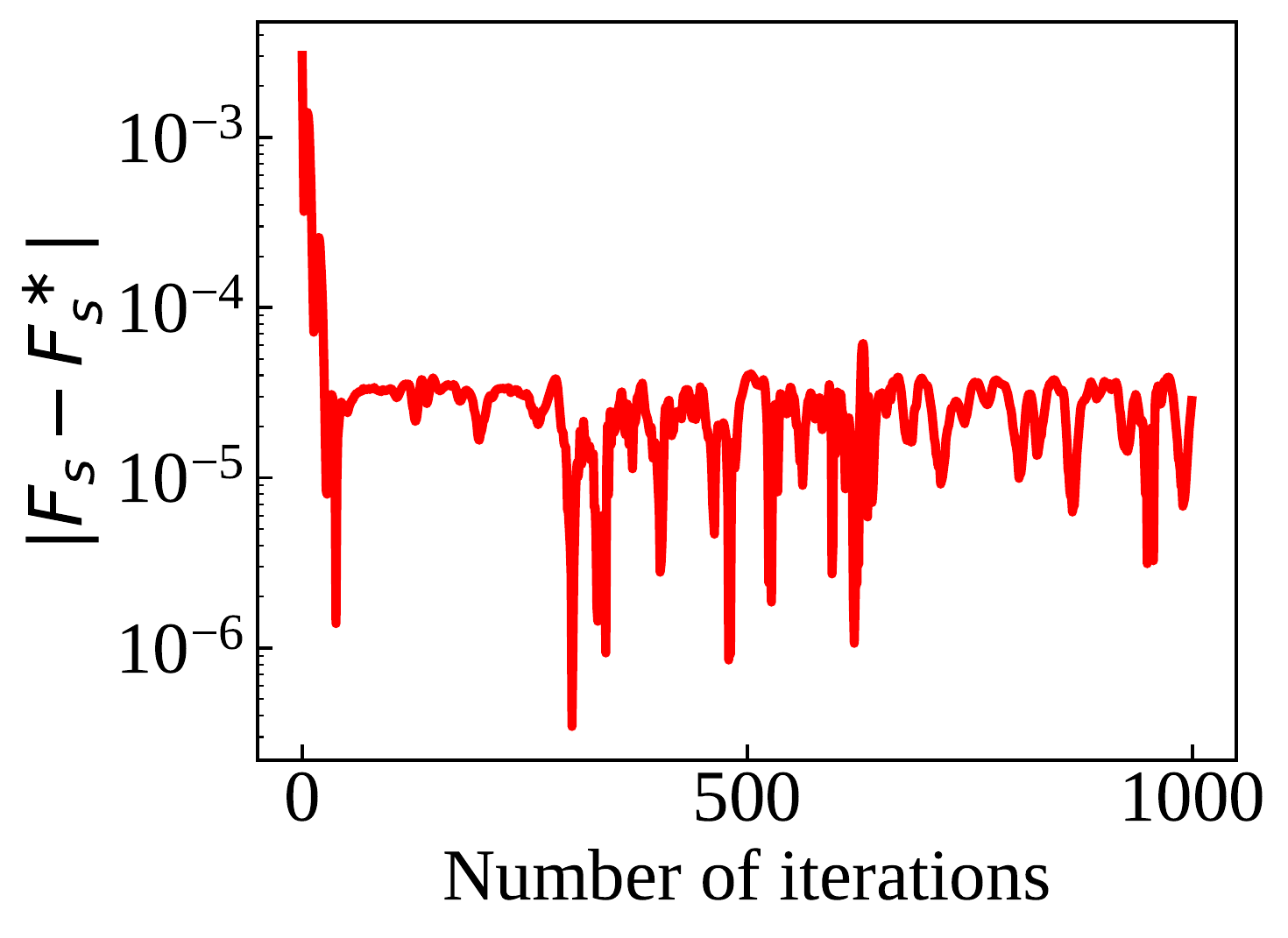} \label{subfig:5edges_shots_history_struct}}
	\caption[]{Result of the five-edge-design problem. Probability distribution of the measurement outcomes of $\Ket{\phi (\bar{\eta})}$ where $\bar{\eta}$ is the optimized parameters obtained by the (a) \textit{statevector simulator} and (b) \textit{QASM simulator}. 
	The bit strings in the horizontal axis represents the structures $\{x_j \}_{j=1}^5$. 
	The history of $|F_u - F_u^\ast|$ obtained by using the (c) \textit{statevector simulator} and the (d) \textit{QASM simulator} where $F_u^\ast$ is the minimum value of $F_u$ in Eq.~\eqref{eq:obj_u}. 
	The history of $|F_s - F_s^\ast|$ obtained by using the (e) \textit{statevector simulator} and the (f) \textit{QASM simulator} where $F_s^\ast$ is the minimum value of $F_s$ in Eq.~\eqref{eq:obj_s}. \label{fig:5edges_result}}
\end{figure}
%%%%%%%%%%%%%%%%%%%%%%%%%%%%%%%%%%%%%%%%%%%%%%%%%%%%
Figures~\ref{fig:5edges_result}\subref{subfig:5edges_probability}, \subref{subfig:5edges_history_state}, and \subref{subfig:5edges_history_struct} illustrates the optimization results of the five-edge-design problem using the \textit{statevector simulator}.
Figure~\ref{fig:5edges_result}\subref{subfig:3edges_probability} shows that the probability concentrates on the structure $11001$, which coincides with the exact optimal structure illustrated in Fig.~\ref{fig:5edges_prob}\subref{subfig:5edges_optstr}.
Figures~\ref{fig:5edges_result}\subref{subfig:3edges_history_state} and \subref{subfig:3edges_history_struct} illustrate the histories of two optimization problems, one of which is for obtaining the response of the system, and the other of which is for amplifying the probability of the optimal structure.
Although some oscillations can be observed in the objective value of $F_u$, both objective function values decreased as the optimization progressed.

\subsubsection*{QASM simulation}
Figures~\ref{fig:5edges_result}\subref{subfig:5edges_shots_probability}, \subref{subfig:5edges_shots_history_state}, and \subref{subfig:5edges_shots_history_struct} illustrates the optimization result obtained using the \textit{QASM simulator}.
As is the case of using the \textit{statevector simulator}, the probability concentrates on the structure $11001$, which coincides with the exact optimal structure, obtained by the brute-force search, shown in Fig.~\ref{fig:5edges_prob}\subref{subfig:5edges_optstr}.
Figures~\ref{fig:5edges_result}\subref{subfig:3edges_shots_history_state} and \subref{subfig:3edges_shots_history_struct} indicate the two optimization histories, one of which is for calculating the response of the system, and the other of which is for amplifying the probability of the optimal structure.
Compared with the case of using the \textit{statevector simulator}, the objective function value of $F_u$ seriously oscillated while it averagely decreased.
This oscillation is due to the statistical error in calculating the objective and its gradient.

%\bibliographystyle{naturemag-doi}
%\bibliography{ref}

\makeatletter\@input{main_aux.tex}\makeatother